\documentclass[pof,twocolumn,amsmath,amssymb,reprint]{revtex4-2}
\usepackage{epsfig}                                                                 
\usepackage{dcolumn}                                                                
\usepackage{bm}                                                                     
\usepackage{xcolor}
\let\boldsymbol\pmb
\usepackage[pdfstartview=FitH,bookmarksopen=true,bookmarksopenlevel=1]{hyperref}    

\begin{document}
\title[Condensation of vapor in a corner formed by two intersecting walls]{Capillary condensation of saturated vapor\\in a corner formed by two intersecting walls}
\author{E. S. Benilov}
 \email[Email address: ]{Eugene.Benilov@ul.ie}
 \homepage[\newline Homepage: ]{https://staff.ul.ie/eugenebenilov/}
 \affiliation{Department of Mathematics and Statistics, University of Limerick, Limerick V94~T9PX, Ireland}

\begin{abstract}
The dynamics of saturated vapor between two intersecting walls is examined. It
is shown that, if the angle $\phi$ between the walls is sufficiently small,
the vapor becomes unstable, and spontaneous condensation occurs in the corner,
similar to the so-called capillary condensation of vapor into a porous medium.
As a result, an ever-growing liquid meniscus develops near the corner. The
diffuse-interface model and the lubrication approximation are used to
demonstrate that the meniscus grows if and only if $\phi+2\theta<\pi$, where
$\theta$ is the contact angle corresponding to the fluid/solid combination
under consideration. This criterion has a simple physical explanation: if it
holds, the meniscus surface is concave -- hence, the Kelvin effect causes
condensation. Once the thickness of the condensate exceeds by an order of
magnitude the characteristic interfacial thickness, the volume of the meniscus
starts to grow linearly with time. If the near-vertex region of the corner is
smoothed, the instability can be triggered off only by finite-size
perturbations, such that include enough liquid to cover the smoothed aria by a
microscopically-thin liquid film.

\end{abstract}
\maketitle

\section{Introduction}

Saturated vapor and liquid are supposed to be in equilibrium -- thus, if a
small amount of the latter is placed in a container filled with the former, no
exchange of mass should occur.

This simple conclusion -- no matter how natural -- is misleading: if the
liquid is placed in a sufficiently acute (or not too obtuse) corner, mass
exchange does occur. This result is obtained in the present paper for a narrow
range of parameters, using an elaborate mathematical model -- but it has a
simple qualitative explanation and, thus, is likely to hold generally.

\begin{figure}[!b]
\begin{center}\includegraphics[width=\columnwidth]{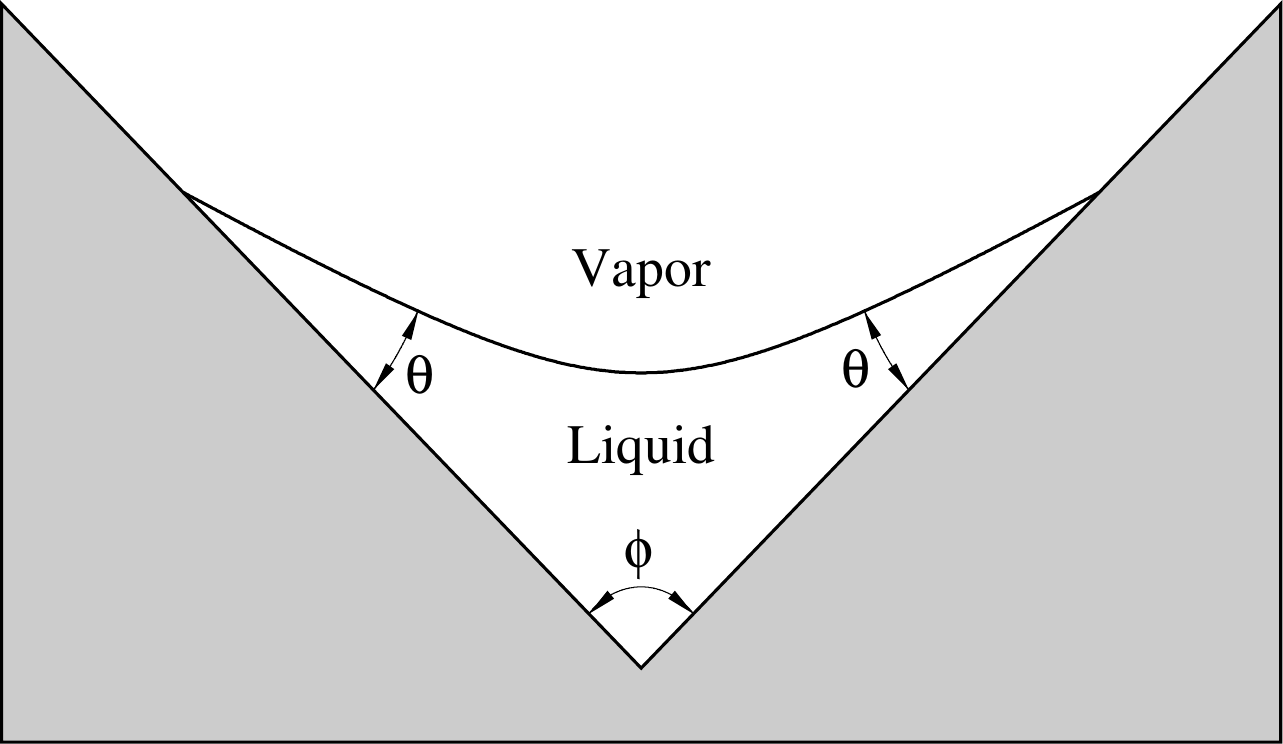}\end{center}
\caption{A liquid meniscus in a corner (the region occupied by the solid is shaded). In the configuration shown, the contact angle $\theta$ is such that $\phi+2\theta <\pi$, so spontaneous condensation occurs.}
\label{fig1}
\end{figure}

Consider a small meniscus in a corner formed by two walls intersecting at an
angle $\phi$ (see Fig. \ref{fig1}), and introduce the microscopic contact
angle $\theta$ at which the meniscus free boundary approaches the walls.
Theoretically, $\theta$ is specific to the fluid/substrate combination under
consideration (e.g., \cite{Davis83}), but in reality the walls are never
perfectly flat and chemically homogeneous. Microscopic imperfections give rise
to a hysteresis interval, i.e., a certain spread in $\theta$ (e.g.,
\cite{SavvaKalliadasis13}); in what follows, it is assumed narrow -- hence,
insignificant -- and is neglected.

Now, let $\phi$ and $\theta$ be such that%
\begin{equation}
\phi+2\theta<\pi, \label{1.1}%
\end{equation}
in which case the free surface of the meniscus is concave (see Fig.
\ref{fig1}). As a result, the Kelvin effect
\cite{EggersPismen10,ColinetRednikov11,RednikovColinet13,Morris14,JanecekDoumencGuerrierNikolayev15,SaxtonVellaWhiteleyOliver17,RednikovColinet17,RednikovColinet19}
gives rise to a vapor-to-liquid mass flux, making the meniscus absorb fluid
from the surrounding vapor and grow -- in a manner, similar to the effect of
capillary condensation of vapor into a porous medium (e.g., \cite{Barnes11}).
If, on the other hand, condition (\ref{1.1}) does not hold, the surface of the
meniscus is convex, and the Kelvin effect makes it dry up. This explains
physically the nonexistence of solutions describing static liquid ridges
\cite{Benilov20c} and three-dimensional drops \cite{Benilov21b} -- in both
cases, on a flat substrate and surrounded by saturated vapor.

The two possible behaviors of menisci could be described using the classical
Navier--Stokes equations, coupled to a model of vapor diffusion in the
surrounding air, with a boundary condition describing condensation and/or
evaporation at the interface (e.g.,
\cite{DeeganBakajinDupontHuberEtal00,DunnWilsonDuffyDavidSefiane09,EggersPismen10,ColinetRednikov11,RednikovColinet13,Morris14,StauberWilsonDuffySefiane14,StauberWilsonDuffySefiane15,JanecekDoumencGuerrierNikolayev15,SaxtonWhiteleyVellaOliver16,SaxtonVellaWhiteleyOliver17,RednikovColinet19,WrayDuffyWilson19}%
). Alternatively (as done in the present paper), the problem can be examined
using the diffuse-interface model: it includes both hydro- and thermodynamics
and, thus, consistently describes all of the effects arising in the problem at hand.

The diffuse-interface model (DIM) was invented as a tool for modeling
interfaces, based on two assumptions put forward by \cite{Korteweg01} in
application to equilibrium interfaces in fluids:

\begin{enumerate}
\item the van der Waals intermolecular force (responsible for phase
transitions) can be described by a pair-wise potential,

\item the characteristic length of this potential is much smaller than the
interfacial thickness.
\end{enumerate}

\noindent In recent times, the DIM was incorporated into non-equilibrium fluid
dynamics (see \cite{AndersonMcfaddenWheeler98,PismenPomeau00} and references
therein) and applied to numerous problems including nucleation and collapse of
bubbles
\cite{MagalettiMarinoCasciola15,MagalettiGalloMarinoCasciola16,GalloMagalettiCasciola18,GalloMagalettiCoccoCasciola20}%
, phase separation in polymer blends
\cite{ThieleMadrugaFrastia07,MadrugaThiele09}, contact lines
\cite{SibleyNoldSavvaKalliadasis14,BorciaBorciaBestehornVarlamovaHoefnerReif19}%
, contact lines in fluids with surfactants
\cite{ZhuKouYaoWuYaoSun19,ZhuKouYaoLiSun20}, Faraday instability
\cite{BorciaBestehorn14,BestehornSharmaBorciaAmiroudine21}, Rayleigh--Taylor instability \cite{ZanellaTegzeLetellierHenry20}, etc. The
DIM was shown to follow from the Enskog--Vlasov kinetic theory
\cite{Giovangigli20,Giovangigli21} -- the same way the usual compressible
hydrodynamics follows from Enskog's theory of dense fluids
\cite{VanbeijerenErnst73a}. \cite{CahnHilliard58} formulated the DIM as a
single equation applicable, under certain conditions, to all systems with
phase transitions and interfaces. An incompressible version of the DIM was
formulated by \cite{JasnowVinals96} and applied to various problems involving
contact lines (e.g. \cite{Jacqmin00,DingSpelt07,YueZhouFeng10,YueFeng11}).

The DIM has been used for modelling settings involving the Kelvin effect.
\cite{Benilov20c,Benilov21b} argued that two- and three-dimensional sessile
drops cannot be static due to the Kelvin-effect-induced evaporation.
\cite{Benilov22a} examined the dynamics of a spherical drop floating in under-
or oversaturated vapor of the same fluid: it was shown that the evaporation in
this case in caused by advection of vapor by an outward flow due to a weak
imbalance between the chemical potentials of the liquid and vapor. In
mixtures, this mechanism acts alongside the diffusion (say, of vapor in air
examined in
\cite{DeeganBakajinDupontHuberEtal00,DunnWilsonDuffyDavidSefiane09,EggersPismen10,ColinetRednikov11,RednikovColinet13,Morris14,StauberWilsonDuffySefiane14,StauberWilsonDuffySefiane15,JanecekDoumencGuerrierNikolayev15,SaxtonWhiteleyVellaOliver16,SaxtonVellaWhiteleyOliver17,RednikovColinet19,WrayDuffyWilson19}%
) -- but in pure fluids (which do not diffuse), it is the \emph{only}
mechanism of evaporation. This makes the DIM an excellent tool for
studying phase transitions in pure fluids.

The present paper applies the original (compressible) version of the DIM to a
pure fluid bounded by two intersecting walls, under an additional assumption
that the angle between the walls is almost straight ($\phi\approx\pi$) and
they are made of a hydrophilic material ($\theta\ll1$). This way, one can
simplify the problem through the lubrication approximation -- and even more
so, since the lubrication approximation for a flat substrate ($\phi=\pi$) is
already in place \cite{Benilov20d}, as is a framework for estimating the DIM
parameters for a specific fluid \cite{Benilov20a}.

In Sec. \ref{Sec 2} of the present paper, the problem will be formulated
mathematically. Secs. \ref{Sec 3}--\ref{Sec 4} examine solutions describing
static and evolving menisci, respectively. The lubrication approximation of
the DIM is derived in Appendix \ref{Appendix B} and summarized in
Sec. \ref{Sec 5} in a self-contained form that can be used for modeling thin
drops with moving contact lines. Sec. \ref{Sec 5} also provides an estimate of
the dimensional timescale of capillary condensation of liquid films for a
real-life example.

\section{Formulation\label{Sec 2}}

\subsection{Thermodynamics}

The thermodynamic properties of a fluid can be described by the dependence of
its internal energy $e$ and entropy $s$ (both specific, or per unit mass) on
the density $\rho$ and temperature $T$ \cite{GiovangigliMatuszewski13}. The
functions $e(\rho,t)$ and $s(\rho,t)$ are not fully arbitrary, as they should
satisfy the fundamental thermodynamic (Gibbs) relation, which can be written
in the form%
\begin{equation}
\frac{\partial e}{\partial T}=T\frac{\partial s}{\partial T}. \label{2.1}%
\end{equation}
Then, the equation of state (the expression for the pressure $p$ as a function
of $\rho$ and $T$) is given by%
\begin{equation}
p=\rho^{2}\left(  \frac{\partial e}{\partial\rho}-T\frac{\partial s}%
{\partial\rho}\right)  , \label{2.2}%
\end{equation}
and the specific chemical potential, or Gibbs free energy, by%
\begin{equation}
G=e-Ts+\frac{p}{\rho}. \label{2.3}%
\end{equation}
It follows from (\ref{2.1})--(\ref{2.3}) that%
\begin{equation}
\frac{\partial p}{\partial\rho}=\rho\dfrac{\partial G}{\partial\rho},
\label{2.4}%
\end{equation}%
\begin{equation}
\dfrac{\partial p}{\partial T}=\rho\left(  \dfrac{\partial G}{\partial
T}+s\right)  , \label{2.5}%
\end{equation}%
\begin{equation}
\dfrac{\partial G}{\partial T}=-\frac{\partial\left(  \rho s\right)
}{\partial\rho}. \label{2.6}%
\end{equation}
These identities will be needed later, as well as the definition of the
parameter%
\[
B=p-\rho^{2}\frac{\partial e}{\partial\rho}.
\]
$B(\rho,T)$ is not one of the standard thermodynamic functions, but it is
convenient when thermodynamics is coupled to fluid dynamics. It characterizes
the production/consumption of thermal energy due to mechanical
compression/expansion of the fluid (more details to follow). Using definition
(\ref{2.2}) of $p$, one can represent $B$ in the form%
\begin{equation}
B=-\rho^{2}T\frac{\partial s}{\partial\rho}. \label{2.7}%
\end{equation}

\subsection{Governing equations}

A flow of a non-ideal fluid can be characterized by the density $\rho$,
temperature $T$, and velocity $\mathbf{v}=\left(  u,v,w\right)  $ -- depending
on the spatial coordinates $\left(  x,y,z\right)  $ and time $t$. Assume also
that the fluid is affected by a bulk force $\mathbf{F}$, which will be later
identified with the intermolecular attraction (sometimes referred to as the
van der Waals attraction).

Using the identity%
\[
\frac{1}{\rho}\boldsymbol{\boldsymbol{\nabla}}%
p=s\boldsymbol{\boldsymbol{\nabla}}T+\boldsymbol{\boldsymbol{\nabla}}G
\]
[which follows from (\ref{2.4})--(\ref{2.5})], one can write the standard
hydrodynamic equations in the form%
\begin{equation}
\frac{\partial\rho}{\partial t}+\boldsymbol{\boldsymbol{\nabla}}\cdot\left(
\rho\mathbf{v}\right)  =0, \label{2.8}%
\end{equation}%
\begin{equation}
\frac{\partial\mathbf{v}}{\partial t}+\left(  \mathbf{v}\cdot
\boldsymbol{\boldsymbol{\nabla}}\right)  \mathbf{v}%
+s\boldsymbol{\boldsymbol{\nabla}}T+\boldsymbol{\boldsymbol{\nabla}}G=\frac
{1}{\rho}\boldsymbol{\boldsymbol{\nabla}}\cdot\boldsymbol{\Pi}+\mathbf{F},
\label{2.9}%
\end{equation}%
\begin{multline}
\rho c\left(  \frac{\partial T}{\partial t}+\mathbf{v}\cdot
\boldsymbol{\boldsymbol{\nabla}}T\right)  +B\boldsymbol{\boldsymbol{\nabla}%
}\cdot\mathbf{v}\\
=\boldsymbol{\Pi}:\boldsymbol{\boldsymbol{\nabla}}\mathbf{v}%
+\boldsymbol{\boldsymbol{\nabla}}\cdot\left(  \kappa
\boldsymbol{\boldsymbol{\nabla}}T\right)  , \label{2.10}%
\end{multline}
where the dotless product of two vectors produces a second-order tensor, the
symbol \textquotedblleft\thinspace$:$\thinspace\textquotedblright\ denotes the
double scalar product of such tensors,%
\begin{equation}
\boldsymbol{\Pi}=\mu_{s}\left[  \boldsymbol{\boldsymbol{\nabla}}%
\mathbf{v}+\left(  \boldsymbol{\boldsymbol{\nabla}}\mathbf{v}\right)
^{T}-\frac{2}{3}\mathbf{I}\left(  \boldsymbol{\boldsymbol{\nabla}}%
\cdot\mathbf{v}\right)  \right]  +\mu_{b}\,\mathbf{I}\left(
\boldsymbol{\boldsymbol{\nabla}}\cdot\mathbf{v}\right)  \label{2.11}%
\end{equation}
is the viscous stress tensor, $\mathbf{I}$ is the identity matrix, $\mu_{s}$
($\mu_{b}$) is the shear (bulk) viscosity, $\kappa$ is the thermal
conductivity, and $c$ is the heat capacity at constant volume (the
traditionally used subscript $_{V}$ is omitted).

Note that $\mu_{s}$, $\mu_{b}$, $\kappa$, $c$, and $B$ depend generally on
$\rho$ and $T$. Observe also that the term involving $B$ in Eq. (\ref{2.10})
describes the production or consumption of thermal energy due to the fluid's
compression ($\boldsymbol{\boldsymbol{\nabla}}\cdot\mathbf{v}<0$) or expansion
($\boldsymbol{\boldsymbol{\nabla}}\cdot\mathbf{v}>0$), respectively.

The diffuse-interface model (DIM) assumes the following expression for the van
der Waals force:%
\begin{equation}
\mathbf{F}=K\boldsymbol{\boldsymbol{\nabla}}\nabla^{2}\rho, \label{2.12}%
\end{equation}
where the Korteweg parameter $K$ is a fluid-specific constant, not depending
on $\rho$ and $T$.

Eqs. (\ref{2.8})--(\ref{2.11}) (with an unspecified force $\mathbf{F}$) have
been derived by \cite{VanbeijerenErnst73a} from Enskog's theory of dense
fluids. For numerous other derivations, through irreversible thermodynamics
and similar models, see the references cited by \cite{Giovangigli20} and
\cite{GiovangigliMatuszewski13}. The full set (\ref{2.8})--(\ref{2.12}),
including the expression for $\mathbf{F}$, was derived by \cite{Giovangigli20}
from the Enskog--Vlasov kinetic equation.

\subsection{Boundary conditions at the substrate}

\begin{figure}
\begin{center}\includegraphics[width=\columnwidth]{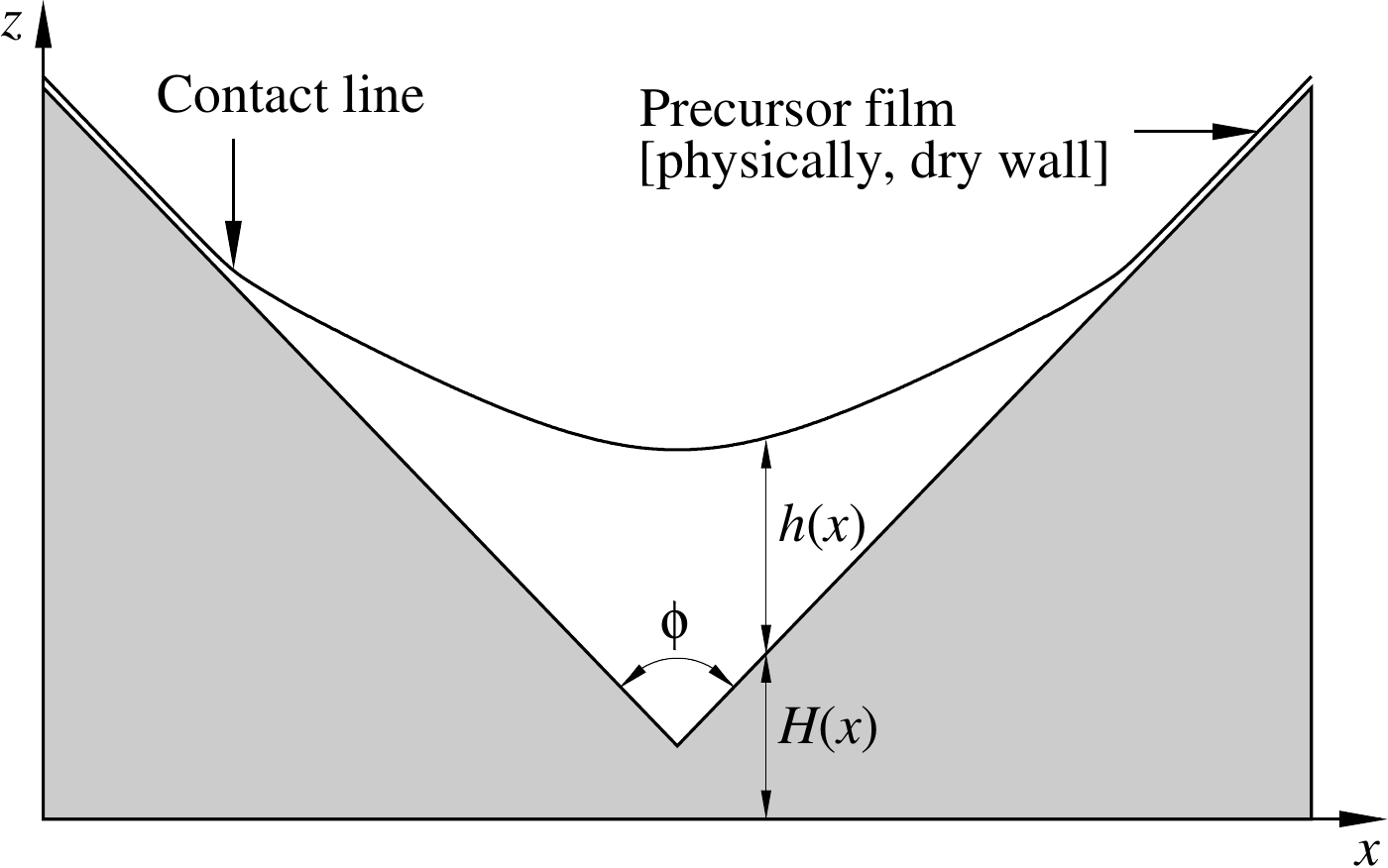}\end{center}
\caption{Formulation of the problem.}
\label{fig2}
\end{figure}

Assume that the fluid is bounded below by a solid substrate whose shape is
given by $z=H(x,y)$ -- see Fig. \ref{fig2}. This implies the no-flow boundary
condition,%
\begin{equation}
\mathbf{v}=\mathbf{0}\qquad\text{at}\qquad z=H.\label{2.13}%
\end{equation}
Let the substrate be kept at a fixed temperature,%
\begin{equation}
T=T_{0}\qquad\text{at}\qquad z=H.\label{2.14}%
\end{equation}
Physically, this boundary condition implies that the substrate is sufficiently
thick, and the heat conductivity of the material it is made of is sufficiently
large -- in which case it is able to `hold' its temperature regardless of the
heat flux coming from the fluid. Note also that the results of this work are
not sensitive to the choice of the boundary condition for the temperature, and so (\ref{2.14}) could be replaced with, say, the condition of
insulation (zero heat flux).

Due to the presence of higher-order derivatives of $\rho$ in expression
(\ref{2.12}) for the van der Waals force, an extra boundary condition is
required for the density. There are several versions of such in the literature
(e.g., \cite{Seppecher96,PismenPomeau00}), of which the simplest one is used
in this work,%
\begin{equation}
\rho=\rho_{0}\qquad\text{at}\qquad z=H, \label{2.15}%
\end{equation}
where $\rho_{0}$ is a phenomenological parameter. The physical meaning of this
condition can be clarified by considering the van der Waals force acting on
the fluid in the near-substrate boundary layer: the solid attracts it
\emph{towards} the substrate, while the fluid outside the boundary layer pulls
it \emph{away} from the substrate. The former force is fixed, whereas the
latter grows with the near-substrate density, so the balance is achieved when
the density assumes a certain value -- which is precisely what condition
(\ref{2.15}) prescribes.

In addition to the advantage of simplicity, condition (\ref{2.15}) can be
derived under the same assumptions as the DIM itself \cite{Benilov20a}.
Furthermore, since the expected effect of spontaneous condensation depends
only on the curvature of the meniscus interface (as argued in the
Introduction), the model used for the boundary condition is not essential.
Condensation occurs at the liquid/vapor interface, so the fluid/substrate
interaction affects it weakly.

\subsection{Boundary conditions far above the substrate}

\begin{figure}
\begin{center}\includegraphics[width=\columnwidth]{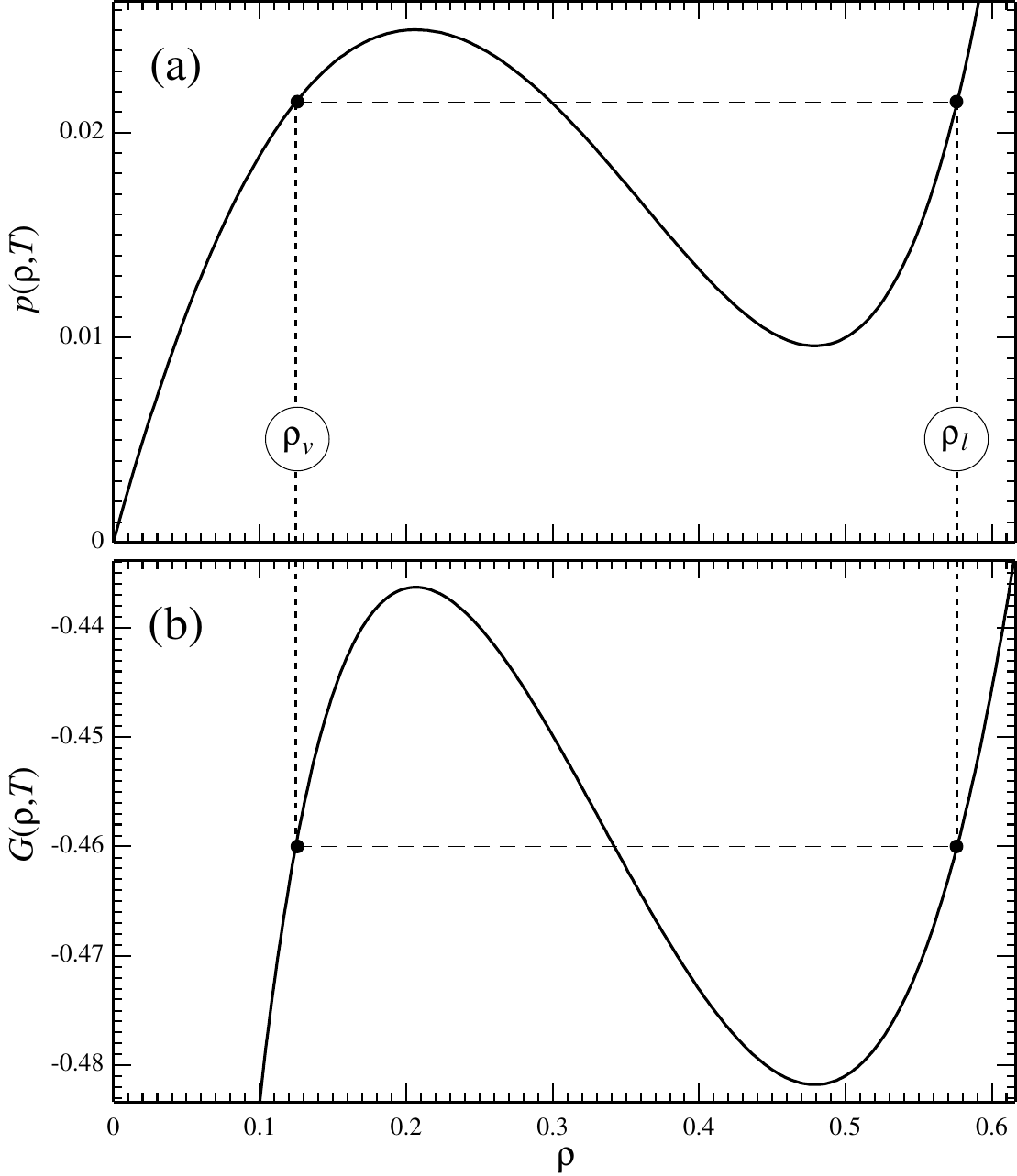}\end{center}
\caption{An illustration of the Maxwell construction. Conditions (\ref{2.19}) and (\ref{2.20}) are illustrated in panels (b) and (a), respectively, for the particular case of the van der Waals fluid (\ref{3.18})--(\ref{3.19}) with $T=0.26$. Observe that $\partial p/\partial\rho$ and $\partial G/\partial\rho$ are positive at both $\rho =\rho_{l}$ and $\rho=\rho_{v}$ (so that the liquid and vapor phases are thermodynamically stable).}
\label{fig3}
\end{figure}

Assume that, far above the substrate, the tangential stress and vertical heat
flux are both zero,%
\begin{equation}
\frac{\partial\mathbf{v}}{\partial z}\rightarrow\mathbf{0}\qquad
\text{as}\qquad z\rightarrow+\infty, \label{2.16}%
\end{equation}%
\begin{equation}
\frac{\partial T}{\partial z}\rightarrow0\qquad\text{as}\qquad z\rightarrow
+\infty. \label{2.17}%
\end{equation}
As explained in the Introduction, this paper is concerned with the dynamics of
\emph{saturated} vapor -- thus, assume%
\begin{equation}
\rho\rightarrow\rho_{v}\qquad\text{as}\qquad z\rightarrow+\infty. \label{2.18}%
\end{equation}
The saturated vapor density $\rho_{v}$, together with the matching liquid
density $\rho_{l}$, depend on the temperature and are determined by the
so-called Maxwell construction,%
\begin{equation}
G(\rho_{v},T)=G(\rho_{l},T), \label{2.19}%
\end{equation}%
\begin{equation}
p(\rho_{v},T)=p(\rho_{l},T). \label{2.20}%
\end{equation}
One should also require that the vapor and liquid be thermodynamically stable,
which amounts to%
\[
\left(  \frac{\partial p}{\partial\rho}\right)  _{\rho=\rho_{v}}\geq
0,\qquad\left(  \frac{\partial p}{\partial\rho}\right)  _{\rho=\rho_{l}}%
\geq0,
\]
i.e., an increase in $\rho$ should not reduce the pressure. Note that $p$ in
the above inequalities can be replaced with the chemical potential $G$, as
their derivatives with respect to $\rho$ are of the same sign [see identity
(\ref{2.4})]. An illustration of the Maxwell construction can be found in Fig.
\ref{fig3}.

For realistic $G(\rho,T)$ and $p(\rho,T)$ and a sufficiently low (\emph{subcritical})
temperature $T$, Eqs. (\ref{2.19})--(\ref{2.20}) admit a unique solution for
the pair $\left(  \rho_{v},\rho_{l}\right)  $ such that $\rho_{v}<\rho_{l}$.
For a sufficiently high (\emph{supercritical}) $T$, (\ref{2.19})--(\ref{2.20}) can
only be satisfied by the trivial solution $\rho_{v}=\rho_{l}$, which
physically means that only one phase exists.  Everywhere in this paper, the temperature is assumed to be subcritical.

Physically, the Maxwell construction ensures that a liquid/vapor interface is
in equilibrium: the equalities of the chemical potential and pressure in the
two phases guarantee the thermodynamic and mechanical equilibria,
respectively. Mathematically, conditions (\ref{2.19})--(\ref{2.20}) can be
derived from the DIM (see below) or any other adequate model by adapting the
governing equations for the static isothermal flat interface in an unbounded space.

Let the near-substrate density prescribed by boundary condition (\ref{2.15})
be such that%
\begin{equation}
\rho_{v}<\rho_{0}<\rho_{l}. \label{2.21}%
\end{equation}
If this condition does not hold, the substrate becomes either perfectly
hydrophobic ($\rho_{0}\leq\rho_{v}$) or perfectly hydrophilic ($\rho_{0}%
\geq\rho_{l}$) \cite{PismenPomeau00,Benilov20a}. In the former case,
condensation cannot occur on the substrate (because it repulses the liquid
phase), whereas the latter implies immediate condensation regardless of all
other parameters.

\subsection{How can liquid and vapor be distinguished in a continuous density
field?}

Since the DIM assumes the density to vary continuously (as opposed to being
restricted to $\rho=\rho_{v}$ or $\rho=\rho_{l}$), one needs a formal
definition of the position of the interface between the phases. The simplest
option is to assume that the fluid with $\rho>\frac{1}{2}\left(  \rho_{v}%
+\rho_{l}\right)  $ should be treated as liquid, and \emph{vice versa}.

Thus, the liquid/vapor interface is defined to be located at the height
$z=H+h$, where $h(x,y,t)$ is such that%
\begin{equation}
\rho(x,y,H+h,t)=\frac{1}{2}\left(  \rho_{v}+\rho_{l}\right)  . \label{2.22}%
\end{equation}
To ensure that $h>0$, one should require [in addition to restriction
(\ref{2.21})] that%
\[
\rho_{0}>\frac{1}{2}\left(  \rho_{v}+\rho_{l}\right)  .
\]
Given this condition, a layer exists adjacent to the substrate, $H<z<H+h$,
which should be mathematically treated as liquid. Most importantly, even if
one considers a horizontally-localized drop or meniscus, this layer stretches
to infinity in all horizontal directions (see a schematic in Fig. \ref{fig2})
-- which was probably what prompted \cite{PismenPomeau00} to dub it a
\textquotedblleft precursor film\textquotedblright. Yet, physically, it
corresponds to dry substrate -- or, equivalently, to the solid/vapor interface.

Let $\bar{h}$ be the thickness of the precursor film on a flat unbounded
substrate; as shown below, $\bar{h}$ depends on the fluid's thermodynamic
properties and the Korteweg parameter $K$. One should keep in mind that, by
comparison with typical sizes of capillary menisci (ranging from
$0.1\,\mathrm{mm}$ to $1\,\mathrm{cm}$), $\bar{h}$ is miniscule (on a
nanoscale). In what follows, such scales will be referred to as
\textquotedblleft microscopic\textquotedblright.

\section{Static menisci\label{Sec 3}}

\subsection{Nondimensionalization\label{Sec 3.1}}

Let the fluid be at rest, $\mathbf{v}=\mathbf{0}$, which also implies
steadiness of the density field, $\partial\rho/\partial t=0$, and
isothermality, $T=T_{0}$ (otherwise the heat flux would generate a flow). With
this in mind, and considering for simplicity the two-dimensional (2D) case,
one can reduce (\ref{2.8})--(\ref{2.12}) to a single equation for $\rho(x,y)$,%
\begin{equation}
K\left(  \frac{\partial^{2}\rho}{\partial x^{2}}+\frac{\partial^{2}\rho
}{\partial z^{2}}\right)  -G(\rho,T)+G(\rho_{v},T)=0, \label{3.1}%
\end{equation}
where $T_{0}$ was re-denoted $T_{0}\rightarrow T$ and the value of the
constant of integration (the last term on the left-hand side) was deduced from
boundary condition (\ref{2.18}). Physically, this (elliptic nonlinear)
equation describes the balance of the van der Waals force and pressure
gradient. For an illustration of the nonlinearity present in Eq. (\ref{3.1})
(via the dependence of $G$ on $\rho$), the reader is referred to fig.
\ref{fig3}(b).

To nondimensionalize Eq. (\ref{3.1}), introduce a characteristic density
$\varrho$, pressure $P$, and the interfacial thickness%
\begin{equation}
l=\sqrt{\frac{K\varrho^{2}}{P}}. \label{3.2}%
\end{equation}
Estimates show that $l$ is on a nanometer scale
\cite{MagalettiGalloMarinoCasciola16,GalloMagalettiCoccoCasciola20,Benilov20a}.

As shown by \cite{PismenPomeau00}, the vertical-to-horizontal aspect ratio of
a liquid film can be identified with%
\[
\varepsilon=\frac{\rho_{l}-\rho_{0}}{\rho_{l}}.
\]
Thus, a meniscus can be regarded thin only if the near-wall density $\rho_{0}$
is close to the liquid density $\rho_{l}$.

The following nondimensional variables will be used:%
\begin{equation}
x_{nd}=\frac{x}{\varepsilon^{-1}l},\qquad z_{nd}=\frac{z}{l},\qquad
H_{nd}=\frac{H}{l}, \label{3.3}%
\end{equation}%
\begin{equation}
\rho_{nd}=\frac{\rho}{\varrho},\qquad T_{nd}=\frac{\varrho RT}{P},
\label{3.3a}%
\end{equation}%
\begin{equation}
p_{nd}=\frac{p}{P},\qquad G_{nd}=\frac{\varrho G}{P}, \label{3.4}%
\end{equation}
where $R$ is the specific gas constant. Introduce also%
\begin{equation}
\left(  \rho_{0}\right)  _{nd}=\frac{\rho_{0}}{\varrho},\qquad\left(  \rho
_{v}\right)  _{nd}=\frac{\rho_{v}}{\varrho},\qquad\left(  \rho_{l}\right)
_{nd}=\frac{\rho_{l}}{\varrho}. \label{3.5}%
\end{equation}
In terms of the new variables, Eq. (\ref{3.1}) and boundary conditions
(\ref{2.15}) and (\ref{2.18}) take the form (the subscript $_{nd}$ omitted)%
\begin{equation}
\varepsilon^{2}\frac{\partial^{2}\rho}{\partial x^{2}}+\frac{\partial^{2}\rho
}{\partial z^{2}}-G(\rho,T)+G(\rho_{v},T)=0, \label{3.6}%
\end{equation}%
\begin{equation}
\rho=\rho_{l}-\varepsilon\qquad\text{at}\qquad z=0, \label{3.7}%
\end{equation}%
\begin{equation}
\rho\rightarrow\rho_{v}\qquad\text{as}\qquad z\rightarrow+\infty. \label{3.8}%
\end{equation}

\subsection{1D solutions of Eq. \ref{3.6})\label{Sec 3.2}}

First consider the solution $\bar{\rho}(z)$ of Eq. (\ref{3.6}) that describes
a flat liquid/vapor interface in an unbounded space (i.e., without a
substrate). For this case, Eq. (\ref{3.6}) and boundary condition (\ref{3.8})
become%
\begin{equation}
\frac{\mathrm{d}^{2}\bar{\rho}}{\mathrm{d}z^{2}}-G(\bar{\rho},T)+G(\rho
_{v},T)=0, \label{3.9}%
\end{equation}%
\begin{equation}
\bar{\rho}\rightarrow\rho_{v}\qquad\text{as}\qquad z\rightarrow+\infty,
\label{3.10}%
\end{equation}
whereas the substrate boundary condition should be replaced with%
\begin{equation}
\bar{\rho}\rightarrow\rho_{l\,}\qquad\text{as}\qquad z\rightarrow-\infty.
\label{3.11}%
\end{equation}
Due to the translational invariance of boundary-value problem (\ref{3.9}%
)--(\ref{3.11}), its solution is not unique. To make it such, require%
\begin{equation}
\bar{\rho}(0)=\frac{1}{2}\left(  \rho_{l}+\rho_{v}\right)  . \label{3.12}%
\end{equation}
For a physically meaningful $G(\rho,T)$, $\bar{\rho}(z)$ is a kink-like
function, decreasing monotonically with increasing $z$.

The boundary-value problem for $\bar{\rho}(z)$ can be used to derive the
Maxwell construction. Its first `half' -- equality (\ref{2.19}) -- can be
derived by considering Eq. (\ref{3.9}) in the limit $z\rightarrow-\infty$.
Equality (\ref{2.20}), in turn, can be obtained by multiplying (\ref{3.9}) by
$\mathrm{d}\bar{\rho}/\mathrm{d}z$ and integrating; taking into account
identity (\ref{2.4}) and fixing the constant of integration via boundary
condition (\ref{3.10}), one obtains%
\begin{multline}
\frac{1}{2}\left(  \frac{\mathrm{d}\bar{\rho}}{\mathrm{d}z}\right)  ^{2}%
-\bar{\rho}\left[  G(\bar{\rho},T)-G(\rho_{v},T)\right] \\
+p(\bar{\rho},T)-p(\rho_{v},T)=0. \label{3.13}%
\end{multline}
Considering this equation in the limit $z\rightarrow-\infty$ and using the
(already proven) equality (\ref{2.19}), one can obtain (\ref{2.20}) as required.

The solution $\bar{\rho}(z)$ of Eq. (\ref{3.13}) subject to boundary condition
(\ref{3.12}) can be readily found in an implicit form,\begin{widetext}%
\begin{equation}
\int_{\frac{1}{2}\left(  \rho_{l}+\rho_{v}\right)  }^{\bar{\rho}}%
\frac{2^{-1/2}\mathrm{d}\rho}{\sqrt{\rho\left[  G(\rho,T)-G(\rho
_{v},T)\right]  -p(\rho,T)+p(\rho_{v},T)}}=-z.\label{3.14}%
\end{equation}
Next, introduce a substrate and let it be flat ($H=\operatorname{const}$). The
solution describing this situation can be expressed in terms of the function
$\bar{\rho}(z)$: shifting it to satisfy the boundary condition at the
substrate, one obtains $\rho=\bar{\rho}(z-H-\bar{h})$ where $\bar{h}$ is such
that%
\begin{equation}
\bar{\rho}(-\bar{h})=\rho_{0}.\label{3.15}%
\end{equation}
Physically, $\rho=\bar{\rho}(z-H-\bar{h})$ describes the precursor film on a
dry substrate located at $z=H$, and $\bar{h}$ is the film's nondimensional
thickness. Substituting (\ref{3.14}) into (\ref{3.15}), one obtains%
\begin{equation}
\bar{h}=\int_{\frac{1}{2}\left(  \rho_{l}+\rho_{v}\right)  }^{\rho
_{l}-\varepsilon}\frac{2^{-1/2}\mathrm{d}\rho}{\sqrt{\rho\left[
G(\rho,T)-G(\rho_{v},T)\right]  -p(\rho,T)+p(\rho_{v},T)}}.\label{3.16}%
\end{equation}
\end{widetext}It can be shown (see Appendix \ref{Appendix A.1}) that $\bar{h}$
is logarithmically large,%
\[
\bar{h}=\frac{\ln\varepsilon^{-1}}{C}+\mathcal{O}(1),
\]
where%
\begin{equation}
C=\left[  \left(  \frac{\partial G}{\partial\rho}\right)  _{\rho=\rho_{l}%
}\right]  ^{1/2} \label{3.17}%
\end{equation}
is real (because the liquid was assumed to be thermodynamically stable --
hence, $\left(  \partial G/\partial\rho\right)  _{\rho=\rho_{l}}>0$).

\subsection{An example: the van der Waals fluid\label{Sec 3.3}}

\begin{figure*}
\begin{center}\includegraphics[width=\textwidth]{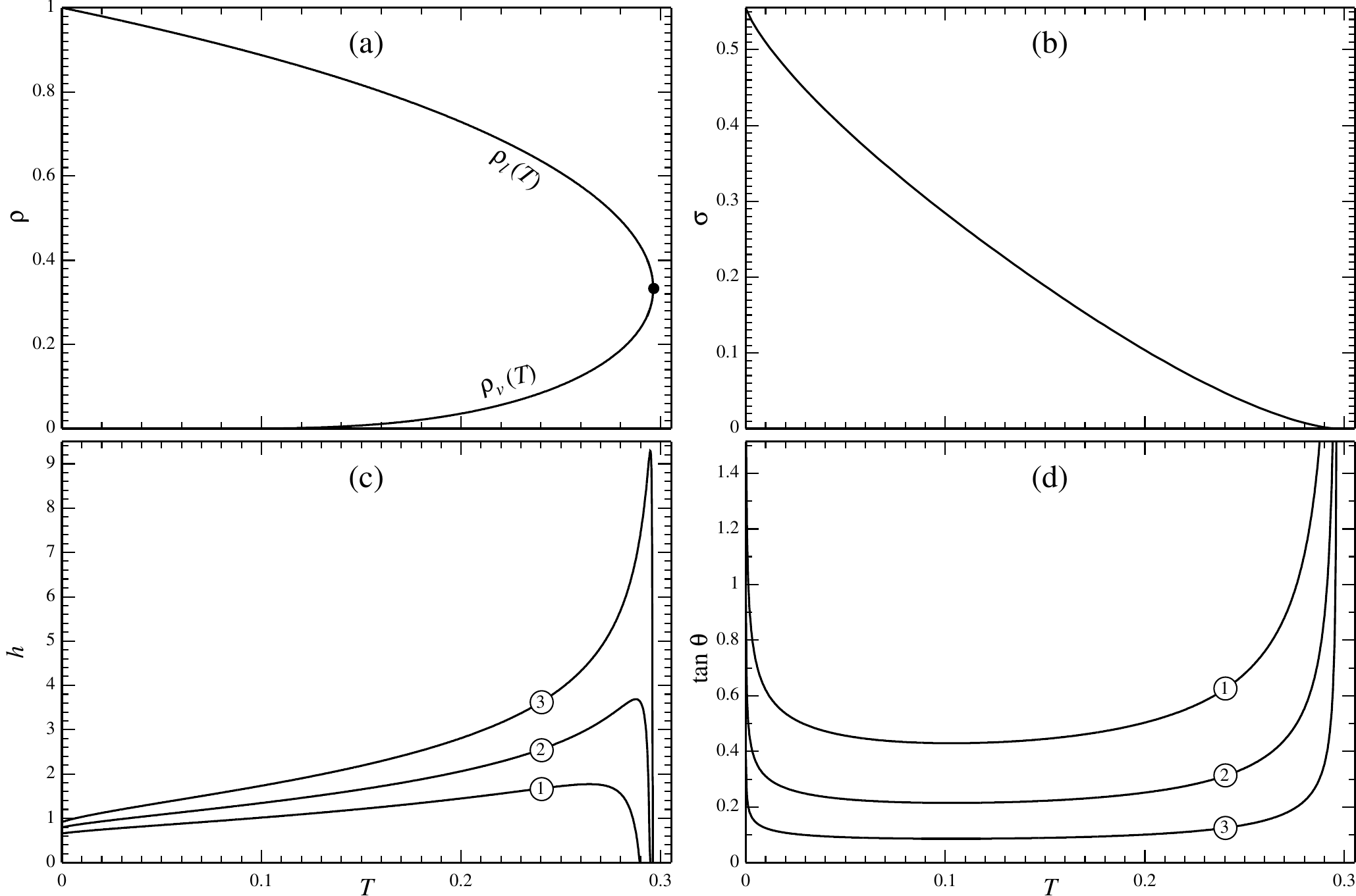}\end{center}
\caption{Various characteristics of interfaces and contact lines \emph{vs} the temperature: (a) densities of the liquid and vapor phases (the black dot marks the critical point); (b) surface tension; (c) precursor film's thickness; (d) $\tan\theta$, where $\theta$ is the contact angle. The curves in panels (c) and (d) correspond to (1) $\varepsilon=0.1$; (2) $\varepsilon =0.05$; (3) $\varepsilon=0.02$.}
\label{fig4}
\end{figure*}

For the van der Waals fluid, the internal energy and entropy (both
nondimensional and specific) are%
\[
e(\rho,T)=cT-\rho,\qquad s(\rho,T)=c\ln T-\ln\frac{\rho}{1-\rho},
\]
where the heat capacity $c$ has been nondimensionalized by the specific gas
constant $R$. The corresponding expressions for pressure (\ref{2.2}) and
chemical potential (\ref{2.3}) are%
\begin{equation}
p(\rho,T)=\frac{T\rho}{1-\rho}-\rho^{2}, \label{3.18}%
\end{equation}%
\begin{multline}
G(\rho,T)=T\left(  \ln\frac{\rho}{1-\rho}+\frac{1}{1-\rho}+c-c\ln T\right) \\
-2\rho+T\left(  1+c-c\ln T\right)  . \label{3.19}%
\end{multline}
The solution of the Maxwell construction (\ref{2.19})--(\ref{2.20}) for this
case is shown in Fig. \ref{fig4}(a) (note that the nondimensional critical
temperature of the van der Waals fluid is $T_{cr}=8/27$).

Note that, for many common fluids at room temperature, $T\lesssim0.1$ (see
Table 1 of \cite{Benilov20b} where $T$ is denoted by $\tau$). Thus, it is
worthwhile to examine the solution of the Maxwell construction in the limit
$T\rightarrow0$. For the van der Waals fluid, it is \cite{Benilov20d}%
\begin{equation}
\rho_{l}=\frac{1+\sqrt{1-4T}}{2}+\mathcal{O}(\operatorname{e}^{-1/T}),
\label{3.20}%
\end{equation}%
\begin{equation}
\rho_{v}=\frac{1+\sqrt{1-4T}}{1-\sqrt{1-4T}}\operatorname{e}^{-1/T}%
+\mathcal{O}(T^{-1}\operatorname{e}^{-2/T}). \label{3.21}%
\end{equation}
Expression (\ref{3.21}) shows that, if $T$ is small, the vapor density is
\emph{exponentially} small, and the same can be assumed for all physically
meaningful equations of states, not only the van der Waals one.

For $T\ll1$, one can deduce from boundary-value problem (\ref{3.9}%
)--(\ref{3.12}) that\begin{widetext}%
\begin{equation}
\bar{\rho}(z)=\left\{
\begin{tabular}
[c]{ll}%
$1+\mathcal{O}(T)\medskip$ & if$\hspace{2.5cm}z\leq-2^{-3/2}\pi,$\\
$\frac{1}{2}\left(  1-\sin2^{1/2}z\right)  +\mathcal{O}(T)\medskip\qquad$ &
if$\qquad-2^{-3/2}\pi\leq z\leq2^{-3/2}\pi,$\\
$0+\mathcal{O}(T)$ & if$\hspace{2.5cm}z\geq2^{-3/2}\pi.$%
\end{tabular}
\ \right.  \label{3.22}%
\end{equation}
\end{widetext}

\subsection{Asymptotic description of static menisci\label{Sec 3.4}}

Consider a static configuration with the liquid phase confined to a layer
adjacent to the substrate, forming a 2D meniscus (liquid film). This implies
that, with increasing $z$, the density first grows from $\rho_{0}$ to
approximately $\rho_{l}$, then decreases towards $\rho_{v}$.

The asymptotic description of menisci with a small aspect ratio is based on
the observation that the general equation (\ref{3.6}) for $\rho(x,z)$ is
asymptotically close to the (much simpler) equation (\ref{3.9}) for $\bar
{\rho}(z)$. Since the difference between the two equations is small, one can
assume%
\begin{equation}
\rho(x,z)\approx\bar{\rho}(z-H-h), \label{3.23}%
\end{equation}
where the undetermined function $h(x)$ is, physically, the distance between
the substrate and fluid/vapor interface (see Fig. \ref{fig2}).

On the basis of assumption (\ref{3.23}), the following asymptotic equation for
$h(x)$ is derived in Appendix \ref{Appendix A}):%
\begin{equation}
\sigma\frac{\mathrm{d}^{2}(H+h)}{\mathrm{d}x^{2}}=f(h-\bar{h}), \label{3.24}%
\end{equation}
where%
\begin{equation}
\sigma=\int_{-\infty}^{\infty}\left(  \frac{\mathrm{d}\bar{\rho}}{\mathrm{d}%
z}\right)  ^{2}\mathrm{d}z \label{3.25}%
\end{equation}
is, physically, the surface tension and the function%
\begin{equation}
f(\xi)=2C^{2}\left(  1-\operatorname{e}^{-C\xi}\right)  \operatorname{e}%
^{-C\xi} \label{3.26}%
\end{equation}
describes the effect exerted on the fluid by the substrate. Recall also that
the precursor film thickness $\bar{h}$ is defined by (\ref{3.16}) and
coefficient $C$, by (\ref{3.17}).

The coefficients $\sigma(T)$ and $\bar{h}(T,\varepsilon)$ have been computed
for the van der Waals fluid [i.e., for $G$ and $p$ given by (\ref{3.18}%
)--(\ref{3.19})] and are shown in Figs. \ref{fig4}(b,c), respectively. The
former figure shows that the surface tension vanishes at the critical point
(as it should). Note also that, since $\rho_{v}\rightarrow\rho_{l}$ as
$T\rightarrow T_{cr}$ [as illustrated in Fig. \ref{fig4}(a)] -- then, sooner
or later, $\rho_{0}=\rho_{l}-\varepsilon$ becomes smaller than $\rho_{v}$.
This violates assumption (\ref{2.21}) and also makes $\bar{h}$ negative, so
this part of the graphs in Fig. \ref{fig4}(c) have been truncated.

Before considering menisci in a corner (which is the ultimate goal of this
paper), it is instructive to examine the solution of Eq. (\ref{3.24}) for a
flat substrate with the following boundary condition:%
\begin{equation}
h\rightarrow\bar{h}\qquad\text{as}\qquad x\rightarrow+\infty. \label{3.27}%
\end{equation}
Substituting $H=\operatorname{const}$ into Eq. (\ref{3.24}), multiplying it by
$\mathrm{d}h/\mathrm{d}z$, integrating with respect to $z$, and fixing the
constant of integration via condition (\ref{3.27}), one can obtain a separable
equation. Its solution will be presented in a form that is best suited for
physical interpretation,%
\begin{equation}
h=\bar{h}+\frac{1}{C}\ln\left[  1+\exp\frac{C\left(  x_{0}-x\right)
\tan\theta}{\varepsilon}\right]  , \label{3.28}%
\end{equation}
where $x_{0}$ is arbitrary and $\theta$ is, at this stage, a constant such
that%
\begin{equation}
\tan\theta=\left(  \frac{2C}{\sigma}\right)  ^{1/2}\varepsilon. \label{3.29}%
\end{equation}
The physical meaning of $\theta$ can be deduced from the asymptotics of
solution (\ref{3.28}) at minus-infinity,%
\[
h\rightarrow-\frac{x}{\varepsilon}\tan\theta\qquad\text{as}\qquad
x\rightarrow-\infty,
\]
which describes a liquid/vapor interface inclined at an angle $\theta$ [the
factor of $1/\varepsilon$ accounts for the different scalings of $x$ and $z$
in nondimensionalization (\ref{3.3})]. Thus, $\theta$ is the contact angle.

The dependence of $\theta$ on $T$, computed for the van der Waals fluid
(\ref{3.18})--(\ref{3.19}) is shown in Fig. \ref{fig4}(d). Observe that
$\tan\theta\rightarrow\infty$ in both small-temperature and near-critical
limits (which can also be deduced analytically from the asymptotic behavior of
$C$ and $\sigma$ as $T\rightarrow0$ and $T\rightarrow T_{cr}$). As a result,
the lubrication approximation fails in these limits, and so the results of
this paper are not applicable.

\begin{figure}
\begin{center}\includegraphics[width=\columnwidth]{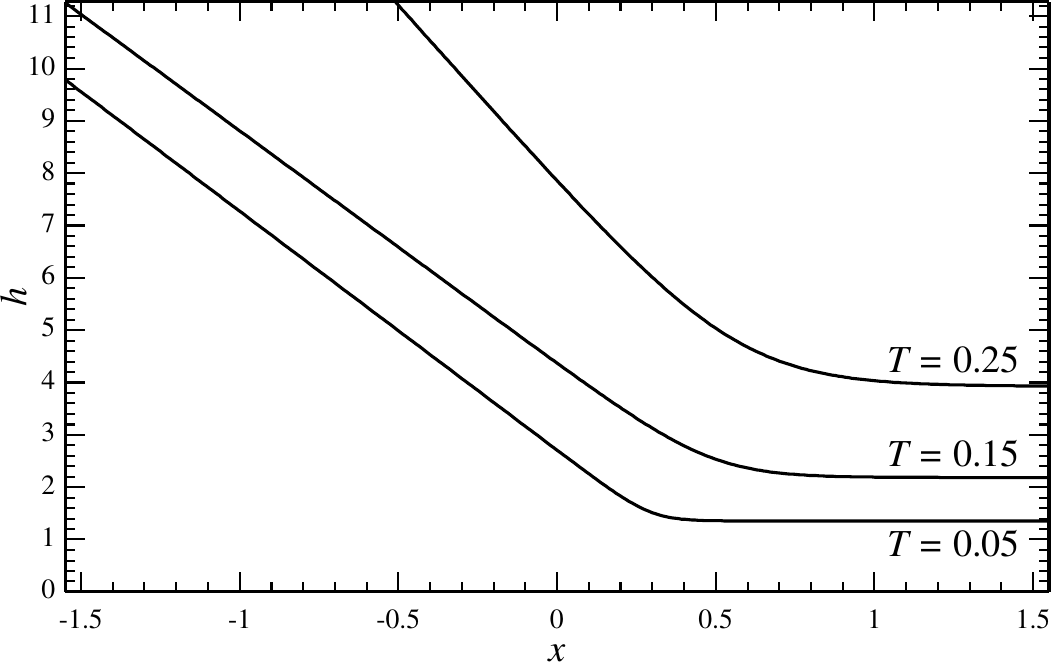}\end{center}
\caption{Examples of solution (\ref{3.28}) for $\varepsilon=0.02$ and three values of the temperature (indicated in the figure).}
\label{fig5}
\end{figure}

Examples of solution (\ref{3.28}), computed for the van der Waals fluid and
various temperatures, are shown in Fig. \ref{fig5}. Observe that the
interfaces for $T=0.05$ and $T=0.15$ are almost parallel, which is a result of
the near-constancy of $\theta$ in the middle part of Fig. \ref{fig4}(d).

\subsection{Static menisci in a corner\label{Sec 3.5}}

Let the substrate form a corner of angle $\phi$ (as in Fig. \ref{fig2}), so
that the substrate is described by%
\begin{equation}
H=\frac{\left\vert x\right\vert }{\varepsilon}\tan\frac{\pi-\phi}{2}.
\label{3.30}%
\end{equation}
Since the lubrication approximation used in this paper implies that $\theta
\ll1$ and $\phi\approx\pi$, \textquotedblleft$\tan$\textquotedblright\ can be
omitted in (\ref{3.29})--(\ref{3.30}), but it can be just as well kept (so
that the results obtained would look more natural).

Given the substrate's symmetry, the meniscus surface should also be symmetric,
which corresponds to the following boundary condition:%
\begin{equation}
\frac{\mathrm{d}(H+h)}{\mathrm{d}x}=0\qquad\text{at}\qquad x=0. \label{3.31}%
\end{equation}
Assume also that, far from the corner the substrate is dry, which implies%
\[
h\rightarrow\bar{h}\qquad\text{as}\qquad x\rightarrow\pm\infty.
\]
Since, in the problem at hand,%
\[
\frac{\mathrm{d}^{2}H}{\mathrm{d}x^{2}}=0\qquad\text{if}\qquad x\neq0,
\]
the general equation (\ref{3.24}) reduces for $x\neq0$ to that for a
\emph{flat} substrate. Using, thus, the same approach, one
obtains\begin{widetext}%
\begin{equation}
h=\bar{h}+\frac{1}{C}\ln\left[  1-\frac{\tan\frac{1}{2}(\pi-\phi)}{\tan
\frac{1}{2}(\pi-\phi)-\tan\theta}\exp\left(  -\frac{C\left\vert x\right\vert
\tan\theta}{\varepsilon}\right)  \right]  .\label{3.32}%
\end{equation}
\end{widetext}Evidently, $h$ is real -- hence, physically meaningful -- only
if%
\begin{equation}
\tfrac{1}{2}\left(  \pi-\phi\right)  \leq\theta. \label{3.33}%
\end{equation}
Not surprisingly, this condition (of existence of static menisci) is the
opposite of condition (\ref{1.1}) of condensation.

Another restriction on the applicability of solution (\ref{3.32}) originates
from the requirement that $h$ be non-negative -- hence,%
\begin{equation}
\tan\tfrac{1}{2}\left(  \pi-\phi\right)  \geq-\left(  \operatorname{e}%
^{C\bar{h}}-1\right)  \tan\theta. \label{3.34}%
\end{equation}
Observe that this condition can fail only if $\phi>180^{\circ}$.

\begin{figure*}
\begin{center}\includegraphics[width=\textwidth]{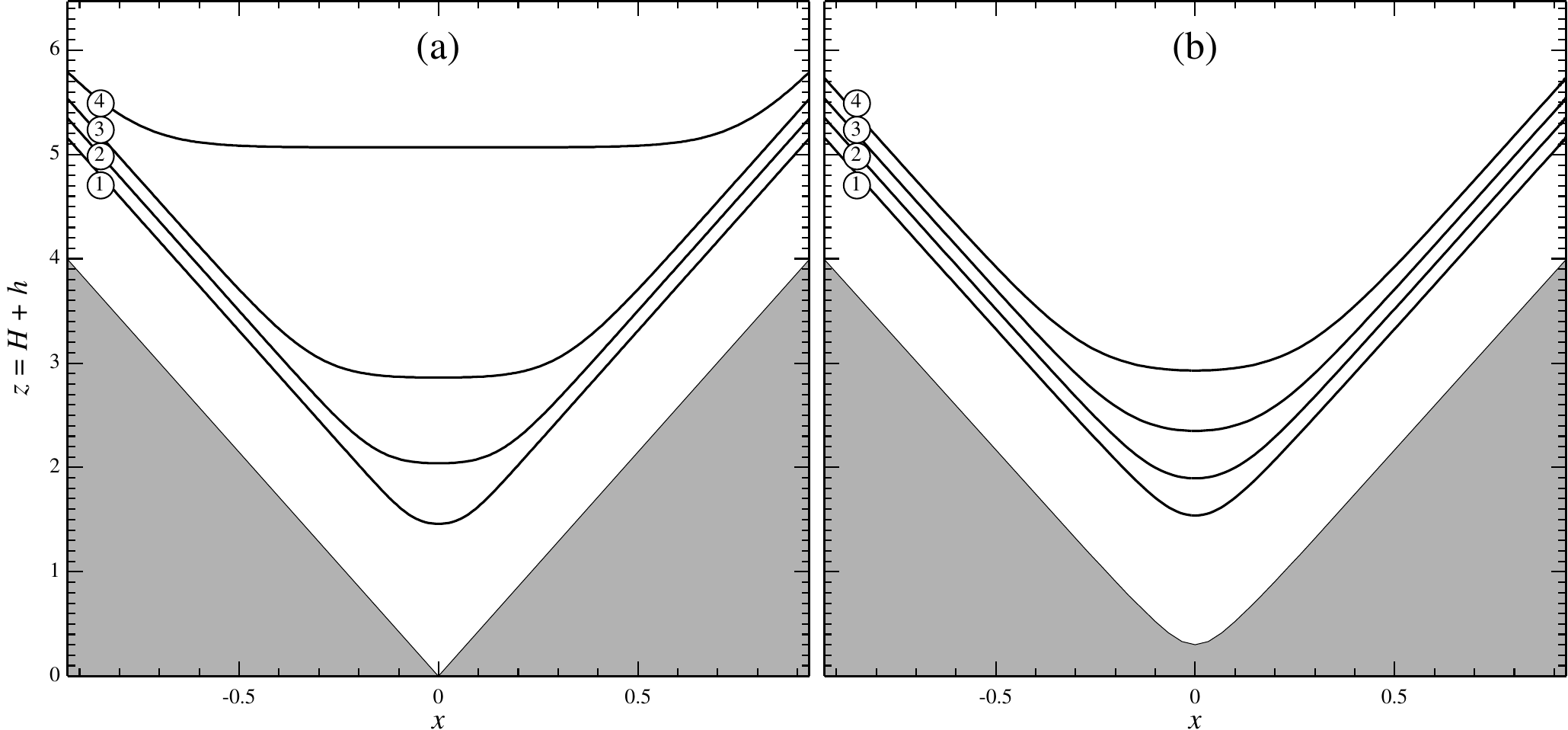}\end{center}
\caption{Examples of static menisci with $\varepsilon=0.02$, in a corner with $\tan\frac{1}{2}\left(  \pi-\phi\right)  =0.085875$. Curves (1)--(4) correspond to $T=0.025,~0.05,~0.075,~0.1$. (a) Solution (\ref{3.32}) for the sharp corner. (b) The numerical solution for the smooth corner described by expression (\ref{3.35})--(\ref{3.36}).}
\label{fig6}
\end{figure*}

Examples of static menisci described by solution (\ref{3.32}) are shown in
Fig. \ref{fig6}(a). They are all computed for the angle $\phi$ such that the
existence condition (\ref{3.33}) holds everywhere except a narrow interval,%
\[
0.1027\lessapprox T\lessapprox0.1033.
\]
Evidently, when $T$ approaches this interval, the core (middle part) of the
meniscus becomes increasingly thick. This does not violate the lubrication
approximation, however, as the slope of the interface remains small.

As for condition (\ref{3.34}), it can be violated -- at least, for the van der
Waals fluid -- only if $T$ is very near its critical value and $\phi$ is near
$2\pi$. These requirements cut out a miniscule part of the problem's parameter
space, not to mention that $\tan\theta$ is not small there -- hence, the
lubrication approximation fails. This effectively means that restriction
(\ref{3.34}) can be ignored.

Note, however, that substrates with a \emph{sharp} corner -- such as the one
given by (\ref{3.30}) -- violate the lubrication approximation. One can still
argue that the corner can be smoothed by an arc with a radius of curvature
much larger than the thickness of the meniscus, but much smaller than the
meniscus's width. In this case, the lubrication approximation holds, yet the
solutions should be asymptotically close to that for the sharp corner.

It turns out, however, that smoothing of the corner changes the nature of the
vapor instability, making this case worth studying. The general tendency will
be illustrated by the following example of the substrate's shape:%
\begin{equation}
H=\sqrt{\left(  \frac{x}{\varepsilon}\tan\frac{\pi-\phi}{2}\right)  ^{2}%
+H_{0}^{2}\operatorname{e}^{-\left(  x/\Delta_{H}\right)  ^{2}}}, \label{3.35}%
\end{equation}
where the constants $H_{0}$ and $\Delta_{H}$ determine the amplitude and width
of smoothing, respectively. In this case, Eq. (\ref{3.24}) cannot be solved
analytically, but its solution can be readily found using the MATLAB function
BVP4c (based on the three-stage Lobatto IIIa formula, see
\cite{KierzenkaShampine01}).

Typical results are shown in Fig. \ref{fig6}(b).
Comparing it with \ref{fig6}(a), one might think that the smoothing reduces
the size of the meniscus -- which is true, but applies mostly to near-critical
menisci (such that $\phi+2\theta\approx\pi$). This occurs because the
smoothing expands their existence region beyond the restriction $\phi+2\theta
\geq\pi$, and so near-critical menisci for a sharp corner are `far-from-critical' for the smoothed one.

Consider, for example, the smoothed corner described by Eq. (\ref{3.35}) with%
\begin{equation}
H_{0}=0.3,\qquad\Delta=1, \label{3.36}%
\end{equation}
in which case numerical computations suggest that menisci exist if
\[
\tan\frac{1}{2}\left(  \pi-\phi\right)  \lesssim0.0872.
\]
For a sharp corner, in turn, the existence condition is given by restriction
(\ref{1.1}) which amounts to%
\[
\tan\frac{1}{2}\left(  \pi-\phi\right)  \leq\tan\theta\approx0.0859.
\]
The difference between the two existence criteria would be too slight to be
important, should it not seem to invalidate the suggested physical
interpretation of the main result of this paper, condition (\ref{1.1}). If
condensation does not occur in a situation where (\ref{1.1}) holds, does this
mean that concave menisci do not absorb moisture?

To resolve the apparent paradox, observe that a sufficiently small drop can
have its contact lines in the \emph{smoothed region} and, thus, not be
sensitive to the global angle $\phi$ -- as a result, it could be static. On
the other hand, a sufficiently large drop with contact lines on the
\emph{flat} parts of the walls should still be unstable.

Mathematically, existence of a static meniscus -- even a stable one with
respect to infinitesimal perturbation -- does not necessarily mean the vapor
is stable with respect to \emph{finite-amplitude} perturbations. This issue
will be clarified in the next section by exploring the meniscus evolution.

\section{Evolving menisci\label{Sec 4}}

As shown in the previous section, steady menisci in a sharp corner exist only
subject to condition (\ref{3.33}), but it remains unclear what happens if
(\ref{3.33}) does not hold. One can only assume that menisci evolve in this case.

To find out how exactly they evolve, two evolution equations, corresponding to
two asymptotic regimes, have been derived: Regime 1 is applicable when
$\rho_{v}\sim\rho_{l}$ (see Appendix \ref{Appendix B.1}) and Regime 2, when
$\rho_{v}\ll\rho_{l}$ (Appendix \ref{Appendix B.2}). According to the former,
the dynamics is dominated by expansion (compression) of the fluid while it
evaporates (condensates) -- whereas, in the latter, these effects are as
strong as advection by horizontal velocity. Motion-induced variations of
temperature are small in both cases, but they can be neglected only in the
latter regime (in the former, they still affect the leading-order dynamics).
Most importantly, Regime 2 applies to many common fluids at room temperature
\cite{Benilov20b} -- and, thus, will be discussed in detail; Regime-1
solutions are qualitatively similar and, thus, will not.

Regime 2 ($\rho_{v}\ll\rho_{l}$) is governed by the following equation:%
\begin{multline}
\frac{\partial h}{\partial t}+\frac{\partial}{\partial x}\left\{
Q(h)\frac{\partial}{\partial x}\left[  \sigma\frac{\partial^{2}(H+h)}{\partial
x^{2}}-f(h-\bar{h})\right]  \right\} \\
=\frac{1}{\varepsilon^{2}A}\left[  \sigma\frac{\partial^{2}(H+h)}{\partial
x^{2}}-f(h-\bar{h})\right]  . \label{4.1}%
\end{multline}
Here, the function $f$ is defined by (\ref{3.26}) and the rest of the
coefficients are%
\begin{equation}
A=0.14219\left[  \mu_{b.v}(T)+\frac{4}{3}\mu_{s.v}(T)\right]  \rho_{l}^{2}%
\rho_{_{v}}^{-5/2}T^{1/2}, \label{4.2}%
\end{equation}%
\begin{equation}
Q(h)=\frac{1}{\rho_{l}^{2}}\int_{0}^{\infty}\frac{\hat{\rho}^{2}(z-h)}{\mu
_{s}(\bar{\rho}(z-h),T)}\mathrm{d}z, \label{4.3}%
\end{equation}
where $\mu_{s.v}(T)=\mu_{s}(\rho_{v},T)$ and $\mu_{b.v}(T)=\mu_{b}(\rho
_{v},T)$ are the shear and bulk viscosities of the vapor, respectively,
$\mu_{s}(\rho,T)$ is the fluid's shear viscosity in the whole density range,
and%
\begin{equation}
\hat{\rho}(z)=\int_{z}^{\infty}\left[  \bar{\rho}(z^{\prime})-\rho_{v}\right]
\mathrm{d}z^{\prime}. \label{4.4}%
\end{equation}
To understand the physical meaning of Eq. (\ref{4.1}), note that the two terms
involving $f$ describe how the substrate affects the liquid/vapor interface
(since $h$ is the distance between the two, it does not come as a surprise
that $f\rightarrow0$ as $h\rightarrow\infty$). Out of the two terms involving
$\sigma$, the one on the left-hand side is the usual capillary term, whereas
the one on the right-hand side describes either evaporation or condensation
due to the Kelvin effect. Which one, depends on the curvature of the
liquid/vapor interface: if it is convex (concave), this term is negative
(positive) and, thus, causes evaporation (condensation). Note also that, if
$H=\operatorname{const}$, Eq. (\ref{4.1}) coincides with its flat-substrate
counterpart derived by \cite{Benilov20d}.

To calculate the function $Q(h)$ [given by (\ref{4.3})--(\ref{4.4})], one
needs to know the shear viscosity $\mu_{s}(\rho,T)$ and chemical potential
$G(\rho,T)$ within the full density range $\rho_{v}<\rho<\rho_{l}$. In this
paper, the simplest approximations are used for these parameters.

To qualitatively model the difference between the shear viscosity of vapor and
that of liquid, it is assumed that%
\begin{equation}
\mu_{s}=\rho, \label{4.5}%
\end{equation}
where the coefficient of proportionality is implied to have been eliminated by
letting the nondimensionalization scale $\mu$ be equal to the shear viscosity
of the liquid phase. Such a choice also makes both $\mu_{s.v}$ and $\mu_{b.v}$ small.

As for $G(\rho,T)$, the van der Waals expression (\ref{3.19}) was used, under
the condition $T\ll1$ (which ensures that $\rho_{v}\ll\rho_{l}$). In this
case, $\bar{\rho}(z)$ is given by expression (\ref{3.22}), and the liquid and
vapor densities, by (\ref{3.20})--(\ref{3.21}) -- or, to leading order,%
\[
\rho_{l}=1,\qquad\rho_{v}=0.
\]
Substituting these equalities, together with (\ref{4.5}) and (\ref{3.22}),
into expressions (\ref{4.3})--(\ref{4.4}), and assuming that $h\geq2^{-3/2}%
\pi\approx1.1107$ (which is not restrictive, as $h$ has already been assumed
to be logarithmically large), one obtains%
\begin{align}
Q(h)  &  =\frac{1}{3}h^{3}+2^{-5/2}\pi\left(  1-\frac{\pi^{2}}{12}\right)
\nonumber\\
&  \approx\frac{1}{3}h^{3}+0.098595. \label{4.6}%
\end{align}
For different $G(\rho,T)$ and $\mu_{s}(\rho,T)$, the numeric factor in the
above formula would be different.

Eq. (\ref{4.1}) requires four boundary conditions: two of these follow from
the symmetry of the problem,%
\begin{equation}
\frac{\partial(H+h)}{\partial x}=\frac{\partial^{3}(H+h)}{\partial x^{3}%
}=0\qquad\text{at}\qquad x=0, \label{4.7}%
\end{equation}
and the others are%
\begin{equation}
h\rightarrow\bar{h},\qquad\frac{\partial h}{\partial x}\rightarrow
0\qquad\text{as}\qquad\left\vert x\right\vert \rightarrow+\infty. \label{4.8}%
\end{equation}

\subsection{Numerical results\label{Sec 4.1}}

Eq. (\ref{4.1}) with its coefficients determined by (\ref{3.16}),
(\ref{3.17}), (\ref{3.25}), (\ref{3.30}), and (\ref{4.6}) was solved
numerically with boundary conditions (\ref{4.7})--(\ref{4.8}), using the
method of lines \cite{Schiesser78}, for numerous initial conditions and in a
wide range of the parameters involved. In all cases where the condensation
criterion (\ref{1.1}) was satisfied, a meniscus grew as $t\rightarrow\infty$.

\begin{figure}
\begin{center}\includegraphics[width=\columnwidth]{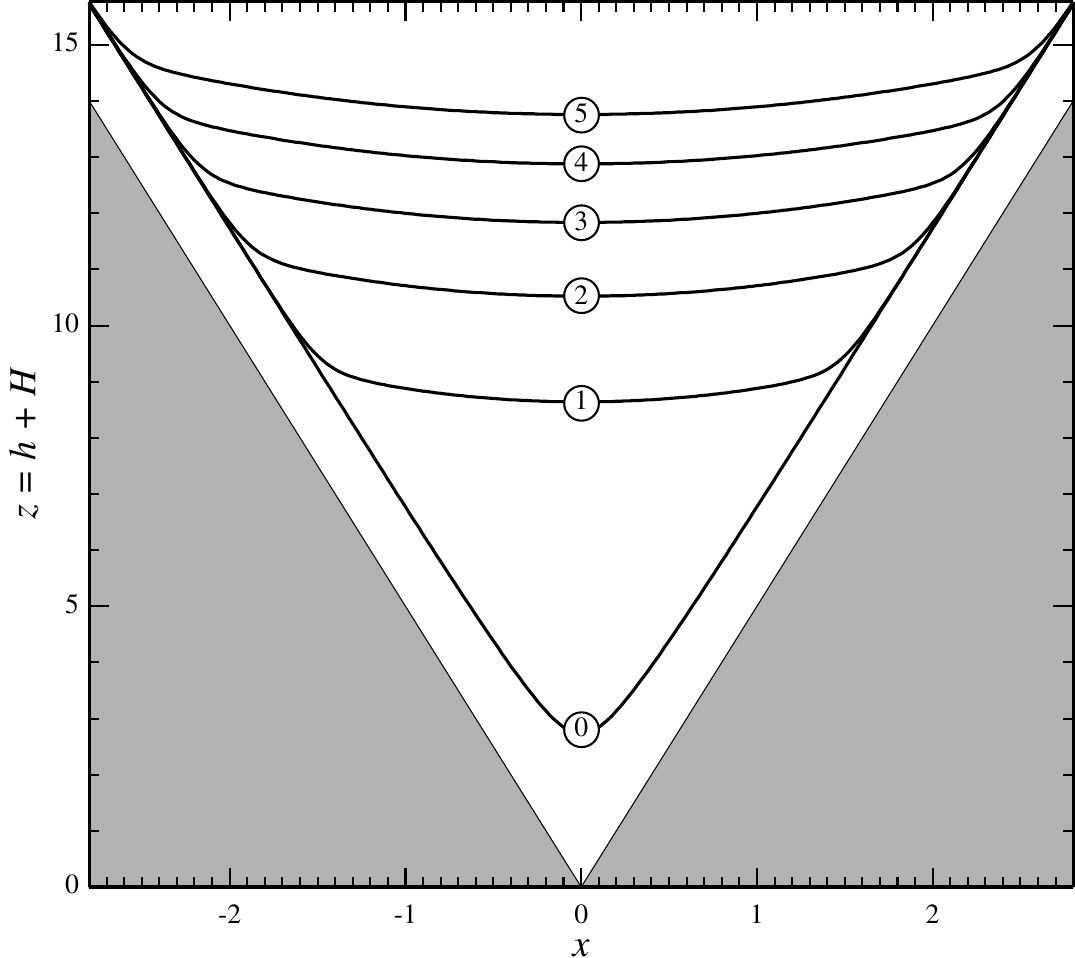}\end{center}
\caption{An example of a meniscus growing in a sharp corner, for parameters (\ref{4.9})--(\ref{4.14}). The curves show `snapshots' of the solution at $t=300\, n$, where $n$ is the curve number (thus, curve $0$ is the initial condition).}
\label{fig7}
\end{figure}

A typical evolution is shown in Fig. \ref{fig7}, computed for the van der
Waals fluid (\ref{3.18})--(\ref{3.19}) with%
\begin{equation}
T=0.1,\qquad\varepsilon=0.02, \label{4.9}%
\end{equation}
in which case the contact angle is $\tan\theta\approx0.0859$. The corner was
such that%
\begin{equation}
\tan\tfrac{1}{2}\left(  \pi-\phi\right)  =0.1, \label{4.10}%
\end{equation}
so that the vapor is weakly unstable. For simplicity, the bulk viscosity of
vapor was assumed to be zero,%
\begin{equation}
\mu_{b.v}=0, \label{4.11}%
\end{equation}
whereas its nondimensional shear viscosity was chosen to match approximately
the ratio of vapor and liquid viscosities of water at room temperature,%
\begin{equation}
\mu_{s.v}=0.01. \label{4.12}%
\end{equation}
The initial condition was\begin{widetext}%
\begin{equation}
H+h=\sqrt{\left(  \frac{x}{\varepsilon}\tan\frac{\pi-\phi}{2}\right)
^{2}+h_{0}\operatorname{e}^{-\left(  x/\Delta_{h}\right)  ^{2}}}%
\qquad\text{at}\qquad t=0,\label{4.13}%
\end{equation}
\end{widetext}with
\begin{equation}
h_{0}=1,\qquad\Delta_{h}=1 \label{4.14}%
\end{equation}
[(\ref{4.13}) looks similar to expression (\ref{3.35}) for $H$, but they are
not to be confused).

The following features of Fig. \ref{fig7} should be observed:

\begin{itemize}
\item[(i)] Initially, a quick adjustment occurs [reflected by the large
difference between curves (0) and (1)].

\item[(ii)] At large times, the growth of the meniscus's thickness and the
progress of its contact lines slow down.
\end{itemize}

\noindent Another feature is quantified in Fig. \ref{fig8} -- which shows the
thickness of the meniscus and the slope of the interface \emph{vs} $x$ for
curve 5 of Fig. \ref{fig7}:

\begin{figure}
\begin{center}\includegraphics[width=\columnwidth]{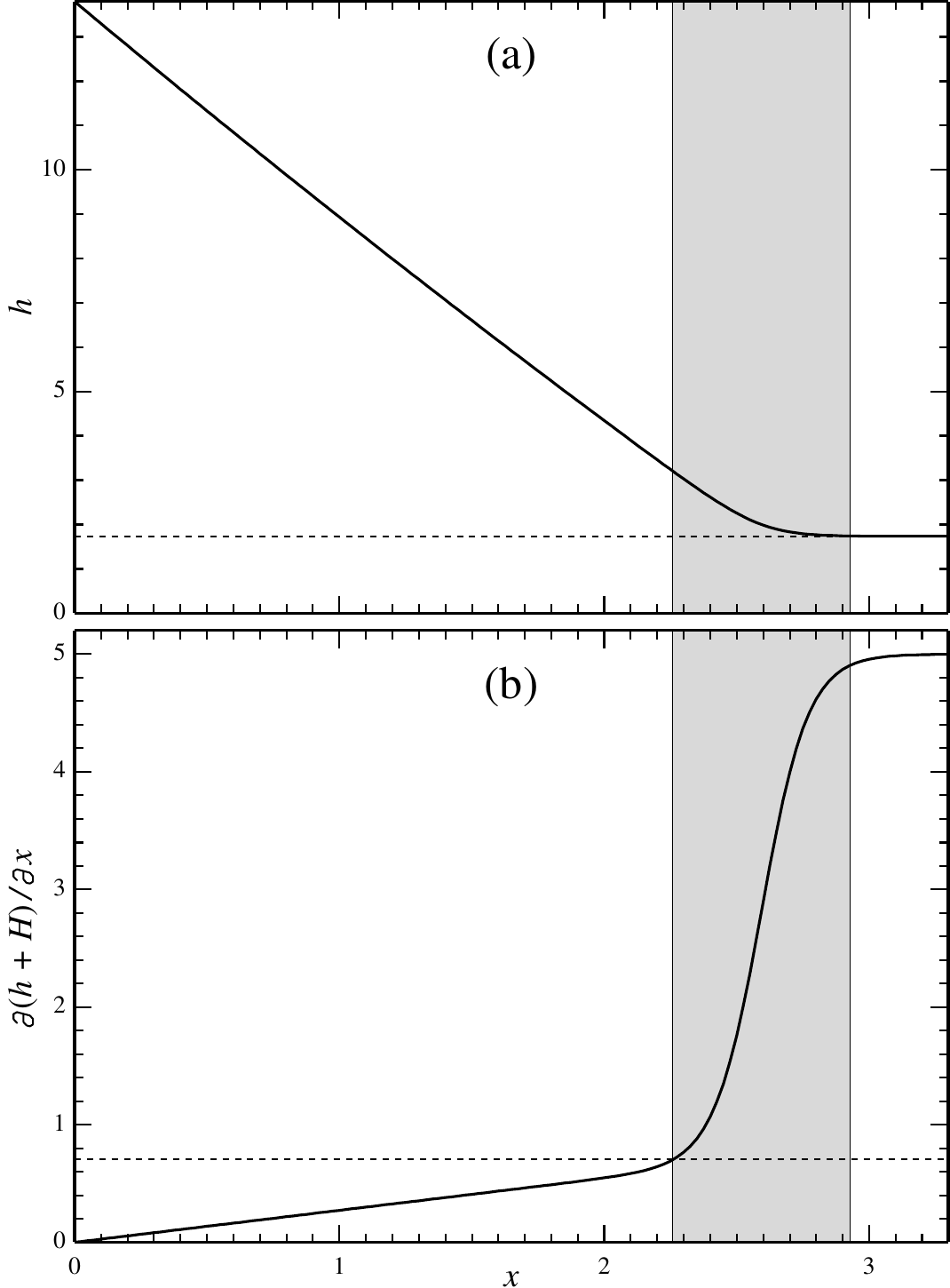}\end{center}
\caption{The cross-section of the growing meniscus in a sharp corner, for parameters (\ref{4.9})--(\ref{4.14}) and $t=1500$ (corresponds to curve 5 in Fig. \ref{fig7}): (a) thickness of the meniscus, $h(x)$; (b) slope of the interface, $\partial (H+h)/\partial x$. The near-contact-line zone is shaded. The dotted line in panel (b) corresponds to the slope given by (\ref{4.15}).}
\label{fig8}
\end{figure}

\begin{itemize}
\item[(iii)] For large times, the `core' of the meniscus assumes a
spherical-cup shape (under lubrication theory, this corresponds to a parabolic
dependence of $h$ on $x$).
\end{itemize}

\noindent Indeed, observe that, in the core, the interfacial slope changes
\emph{linearly} from $0$ (horizontal interface) to%
\begin{equation}
\frac{\partial(H+h)}{\partial x}=\frac{1}{\varepsilon}\left[  \tan\tfrac{1}%
{2}\left(  \pi-\phi\right)  -\tan\theta\right]  \label{4.15}%
\end{equation}
(the angle between the interface and substrate equals $\theta$). As seen in
Fig. \ref{fig8}(b), $\partial(H+h)/\partial x$ assumes value (\ref{4.15}) on
the boundary separating the core and the near-contact-line zone. In the
latter, the thickness of the meniscus is close to that of the precursor film.

\begin{figure}[b]
\begin{center}\includegraphics[width=\columnwidth]{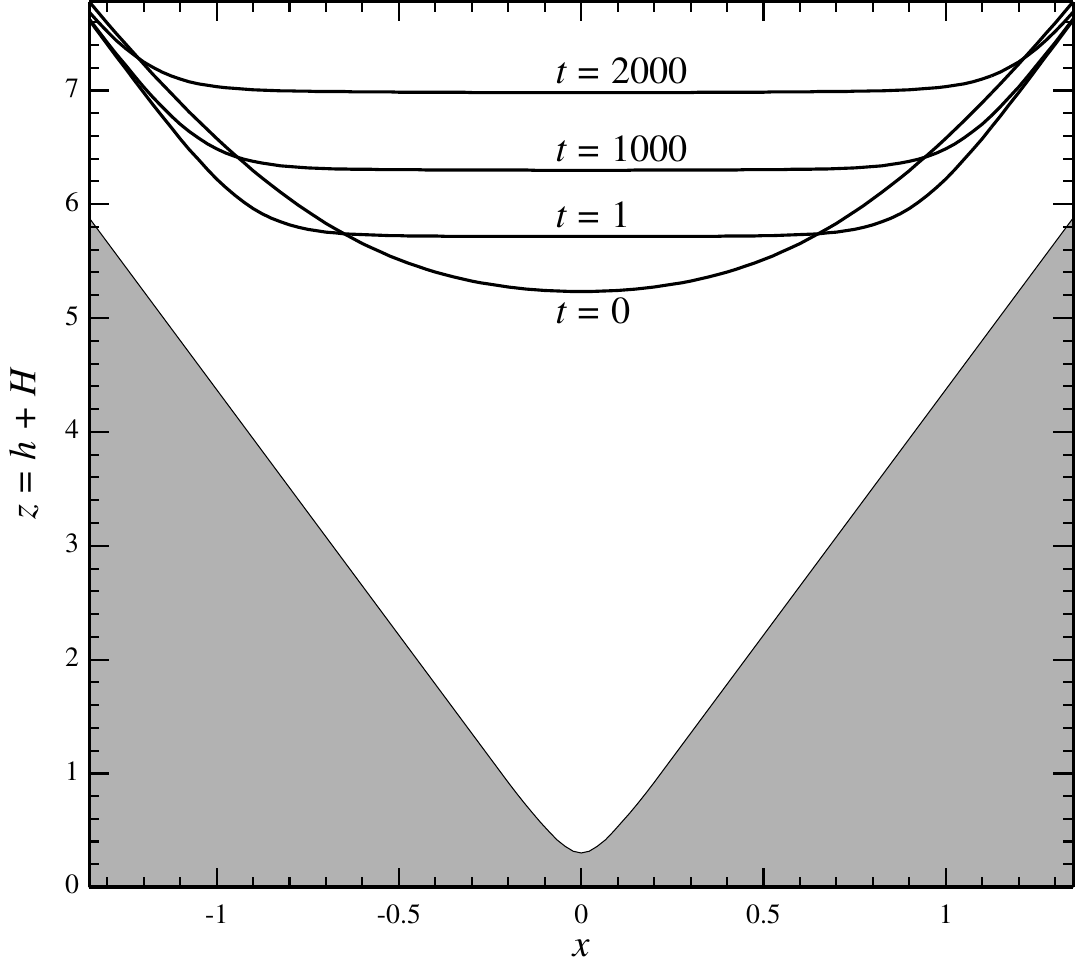}\end{center}
\caption{An example of a growing meniscus in a smooth corner, for parameters (\ref{3.35})--(\ref{3.36}), (\ref{4.9}), (\ref{4.12})--\ref{4.13}), (\ref{4.16})--(\ref{4.17}). The times corresponding to the curves are shown in the figure (observe that they are not equally spaced).}
\label{fig9}
\end{figure}

Fig. \ref{fig9} shows typical evolution of a meniscus in a smoothed corner,
computed for $H(x)$ given by (\ref{3.35})--(\ref{3.36}) with%
\begin{equation}
\tan\frac{\pi-\phi}{2}=0.0871. \label{4.16}%
\end{equation}
The fluid parameters are given by (\ref{4.9}) and the initial condition, by
(\ref{4.13}) with%
\begin{equation}
h_{0}=3.5,\qquad\Delta_{h}=0.5. \label{4.17}%
\end{equation}
Even though a steady solution exists in this case, the meniscus grows as
$t\rightarrow\infty$. It would \emph{not} grow and become static, only if the
amplitude of the initial perturbation is sufficiently small -- e.g., if
$h_{0}=3.5$ in perturbation (\ref{4.17}) is replaced with $h_{0}%
\lessapprox2.9$. One should keep in mind, however, that, since $h$ is
nondimensionbalized by the interfacial thickness $l$, both these values of
$h_{0}$ should be regarded microscopic.

Extensive numerical experiments with various initial conditions showed that,
to make the meniscus grow, its initial volume has to be sufficiently high, but
its shape is unimportant: if it is too `narrow' or too `wide', it spreads out
or contracts, respectively. The end result of the adjustment is a meniscus
with almost flat surface, with its further evolution depending on how wide it
is. For growth, it should cover an area comparable to the smoothed part of the
corner -- otherwise it tends to the existing steady state and becomes static.

Most importantly, the adjusted meniscus does not have to be thick to initiate
growth; its nondimensional thickness can be order-one. In dimensional terms,
this means that the instability is triggered off by \emph{microscopic}
perturbations, i.e., those representing a liquid film whose thickness is
comparable to the interfacial thickness.

As for the large-time evolution of menisci in a smoothed corner, it is
qualitatively the same as that of their sharp-corner counterparts -- i.e.,
both kinds of menisci demonstrate features (ii)--(iii).

\subsection{The large-time behavior\label{Sec 4.2}}

Features (ii)--(iii) of the meniscus evolution listed in the previous
subsection allow one to deduce a simple asymptotic description of the
large-time evolution.

Indeed, feature (iii) suggests that, as $t\rightarrow\infty$, the `outer'
solution (in the meniscus core) is parabolic,%
\begin{equation}
h-\bar{h}\sim h_{0}-h_{2}x^{2}\qquad\text{if}\qquad x_{cl}-x\gg1, \label{4.18}%
\end{equation}
where $h_{0}(t)$ and $h_{2}(t)$ are undetermined functions, and $x_{cl}(t)$ is
the approximate coordinate of the (right-hand) contact line -- so that%
\begin{equation}
h_{0}-h_{2}x_{cl}^{2}=0. \label{4.19}%
\end{equation}
Since, at large times, the meniscus is thick and the contact line is far from
the origin, one should assume $h_{0}\gg1$ and$\ x_{cl}\gg1$, respectively.

According to feature (ii), the velocity of the contact line tends to zero with
time -- hence, the `inner' solution is close to that describing a
\emph{static} contact line. The latter is given by expression (\ref{3.28});
setting in it $x_{0}=x_{cl}+x_{00}$ (where $x_{00}$ is an order-one constant),
one obtains\begin{widetext}%
\begin{equation}
h-\bar{h}\sim\frac{1}{C}\ln\left[  1+\exp\frac{C\left(  x_{cl}+x_{00}%
-x\right)  \tan\theta}{\varepsilon}\right]  \qquad\text{if}\qquad x_{cl}%
-x\sim1.\label{4.20}%
\end{equation}
\end{widetext}where the (order-one) constant $x_{00}$ can only be found from
higher-order approximations.

Matching the outer solution (\ref{4.18}) to the inner solution (\ref{4.20})
effectively amounts to matching their `slopes'; recalling then equality
(\ref{4.19}), one obtains%
\begin{equation}
x_{cl}=\frac{2\varepsilon h_{0}}{\tan\theta},\qquad h_{2}=\frac{\tan^{2}%
\theta}{4\varepsilon^{2}h_{0}}. \label{4.21}%
\end{equation}
It still remains to find $h_{0}(t)$ -- which can be done by either examining
the higher order approximations of the inner and outer solutions -- or,
alternatively, through a simple shortcut involving the exact equation
(\ref{4.1}). To do the latter, integrate (\ref{4.1}) with respect to $x$ from
$0$ to $\infty$ and, recalling boundary conditions (\ref{4.7})--(\ref{4.8})
and the fact that%
\begin{equation}
\frac{\mathrm{d}H}{\mathrm{d}x}\rightarrow\frac{1}{\varepsilon}\tan\frac
{\pi-\phi}{2}\qquad\text{as}\qquad x\rightarrow\infty, \label{4.22}%
\end{equation}
obtain%
\begin{equation}
\frac{\mathrm{d}M}{\mathrm{d}t}=\frac{1}{\varepsilon^{2}A}\left[  \frac
{\sigma}{\varepsilon}\tan\frac{\pi-\phi}{2}-\int_{0}^{\infty}f(h-\bar
{h})\mathrm{d}x\right]  , \label{4.23}%
\end{equation}
where%
\begin{equation}
M=\int_{0}^{\infty}\left(  h-\bar{h}\right)  \mathrm{d}x \label{4.24}%
\end{equation}
is the half-area of the meniscus cross-section. Since the near-contact-line
region is small, $M$ can be estimated using the outer solution (\ref{4.18}).
Observe also that the function $f(h-\bar{h})$ [defined by (\ref{3.26})] is
exponentially small in the outer region -- hence, the integral on the
right-hand side of equality (\ref{4.23}) can be estimated using the inner
solution (\ref{4.20}). Carrying out both estimates and recalling equalities
(\ref{4.21}), one can reduce (\ref{4.23})--(\ref{4.24}) to%
\begin{equation}
\frac{\mathrm{d}M}{\mathrm{d}t}\rightarrow\frac{\sigma}{A\varepsilon^{3}%
}\left(  \tan\dfrac{\pi-\phi}{2}-\tan\theta\right)  \qquad\text{as}\qquad
t\rightarrow\infty, \label{4.25}%
\end{equation}%
\begin{equation}
M\sim\frac{4\varepsilon h_{0}^{2}}{3\tan\theta}\qquad\text{as}\qquad
t\rightarrow\infty. \label{4.26}%
\end{equation}
Thus, the thickness $h_{0}$ of the meniscus grows as $t^{1/2}$, as does its
width $x_{cl}$ [due to (\ref{4.21})] -- whereas the meniscus cross-sectional
area grows linearly. It is also evident from equality (\ref{4.25}) that the
meniscus grows only subject to condition (\ref{1.1}). It should be emphasized
that asymptotic expressions (\ref{4.25})--(\ref{4.26}) hold for both sharp and
smoothed corners [if the latter satisfy condition (\ref{4.22})].

\begin{figure}[b]
\begin{center}\includegraphics[width=\columnwidth]{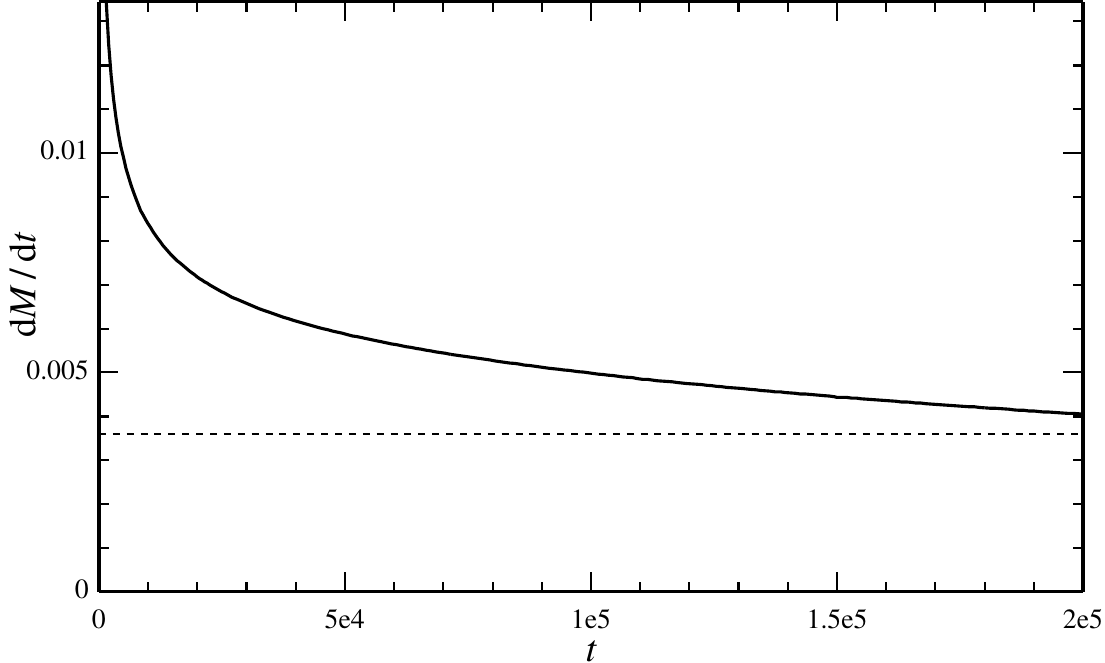}\end{center}
\caption{The long-time evolution of a meniscus with parameters (\ref{4.9})--(\ref{4.14}) (the same as in Fig. \ref{fig7}). $M$ is the half-area of the cross-section [see expression (\ref{4.24})], $t$ is the time. The horizontal dotted line represents the asymptotic value of $\mathrm{d} M/\mathrm{d}t$ predicted by expression (\ref{4.25}).}
\label{fig10}
\end{figure}

To test asymptotics (\ref{4.25}), the exact equation (\ref{4.1}) was solved
numerically for large times. Typical results are illustrated in Fig.
\ref{fig10}: one can see that the rate of growth of the cross-sectional area
of the meniscus does converge to the predicted constant, albeit fairly slowly.

\section{Characteristic time of the Kelvin effect\label{Sec 5}}

To estimate how quick the Kelvin effect is, one needs to rewrite Eq.
(\ref{4.1}) in terms of the dimensional variables and in a form minimizing the
dependence on the fluid's thermodynamic properties (which may not be known in
applications). The DIM parameters -- the Korteweg constant $K$ and the
near-substrate density $\rho_{0}$ -- can be expressed through the surface
tension and contact angle, respectively (for more details, see
\cite{Benilov20a}). The low-temperature assumption $T\ll1$ will also be used,
as it is applicable to many common fluids (including water) at room
temperature \cite{Benilov20b}, and it is also a precondition that $\rho_{v}%
\ll\rho_{l}$, which is required for Eq. (\ref{4.1}) to hold.

Thus, reversing nondimensionalization (\ref{3.3})--(\ref{3.5}), (\ref{B.1}%
)--(\ref{B.3}), (\ref{B.26}) and retaining the same notation for the
dimensional variables, one can write the three-dimensional analogue of Eq.
(\ref{4.1}) in the form\begin{widetext}%
\begin{multline}
\frac{\partial h}{\partial t}+\sigma\mathbf{\nabla}\cdot\left\{
Q(h)\,\mathbf{\nabla}\left[  \nabla^{2}(H+h)-\frac{\tan^{2}\theta}{l}f\left(
\frac{h-\bar{h}}{l}\right)  \right]  \right\}  \\
=\frac{\sigma}{0.14219\left(  \mu_{b.v}+\frac{4}{3}\mu_{s.v}\right)  }\left(
\dfrac{\rho_{v}}{\rho_{l}}\right)  ^{5/2}\left(  \dfrac{K\rho_{l}}{RT}\right)
^{1/2}\left[  \nabla^{2}(H+h)-\frac{\tan^{2}\theta}{l}f\left(  \frac{h-\bar
{h}}{l}\right)  \right]  .\label{5.1}%
\end{multline}
The fluid parameters which appear in this equation and their typical values
are listed in table \ref{table1}. The function%
\[
f(\xi)=\left(  1-\operatorname{e}^{-\xi}\right)  \operatorname{e}^{-\xi}%
\]
is universal (does not involve any parameters), whereas the thickness of the
precursor film $\bar{h}$ and interfacial thickness $l$ depend on the fluid's
chemical potential $G(\rho,T)$ and pressure $p(\rho,T)$,%
\[
l=K^{1/2}\left\{  \left[  \frac{\partial G(\rho,T)}{\partial\rho}\right]
_{\rho=\rho_{l}}\right\}  ^{-1/2},\qquad\bar{h}=K^{1/2}\int_{\frac{1}%
{2}\left(  \rho_{l}+\rho_{v}\right)  }^{\rho_{0}}\frac{2^{-1/2}\mathrm{d}\rho
}{\sqrt{\rho\left[  G(\rho,T)-G(\rho_{v},T)\right]  -p(\rho,T)+p(\rho_{v},T)}%
}.
\]
\end{widetext}The coefficient $Q(h)$ can be expressed through the function
$\bar{\rho}(z)$ describing a flat liquid/vapor interface in an unbounded
space; $\bar{\rho}(z)$, in turn, is related to $G(\rho,T)$ through the
following boundary-value problem:%
\[
K\frac{\mathrm{d}^{2}\bar{\rho}}{\mathrm{d}z^{2}}-G(\bar{\rho},T)+G(\rho
_{v},T)=0,
\]%
\begin{align*}
\bar{\rho}(z)  &  \rightarrow\rho_{l}\,\qquad\text{as}\qquad z\rightarrow
-\infty,\\
\bar{\rho}(z)  &  \rightarrow\rho_{v}\qquad\text{as}\qquad z\rightarrow
+\infty,
\end{align*}%
\[
\bar{\rho}(0)=\frac{1}{2}\left(  \rho_{l}+\rho_{v}\right)  .
\]
Once $\bar{\rho}(x)$ is computed, $Q(h)$ is given by%
\[
Q(h)=\frac{1}{\rho_{l}^{2}}\int_{0}^{\infty}\frac{\hat{\rho}^{2}(z-h)}{\mu
_{s}(\bar{\rho}(z-h),T)}\mathrm{d}z,
\]
where $\mu_{s}(\rho,T)$ is the fluid's shear viscosity, and%
\[
\hat{\rho}(z)=\int_{z}^{\infty}\left[  \bar{\rho}(z^{\prime})-\rho_{v}\right]
\mathrm{d}z^{\prime}.
\]
To calculate $\bar{h}$, $l$, and $Q(h)$, one needs (typically, empiric)
approximations of $G(\rho,T)$ and $p(\rho,T)$, which may not be available for
the liquid used in a specific experiment (say, a certain type of silicone
oil). Even for water -- whose thermodynamic properties are well known -- there
is a problem ensuing from the dependence of $l$ on the derivative $\partial
G/\partial\rho$: even if $G(\rho,T)$ itself is approximated well, its
derivative can be inaccurate (according to the experience of the author of the
present paper).

\renewcommand{\arraystretch}{1.15}
\begin{table*}
\caption{The fluid parameters involved in Eq. \ref{5.1}) and their typical values (at $T=25^{\circ}\mathrm{C}$ and/or $p=1\,\mathrm{atm}$, if appropriate). The values of the parameters not related to the DIM have been borrowed from Ref. \cite{LindstromMallard97,CzerniaSzyk21,HolmesParkerPovey11,ShangWuWangYangYeHuTaoHe19}, and the estimate of $K$, from Ref. \cite{Benilov20a}.} \label{table1}
\begin{ruledtabular}
\begin{tabular}{lll}
Notation & Parameter & Value\vspace*{1mm} \\
\hline
$\rho_{l}$ & density (liquid) & $997.00\,\mathrm{kg\,m}^{-3}$ (water)\vspace*{0.2cm}\\

$\rho_{v}$ & density (vapor)
    & \hspace*{-0.2cm}
      \begin{tabular}[c]{l}
      $0.0231\,\mathrm{kg\,m}^{-3}$ (water)\vspace*{-0.1cm}\\
      $1.1839\,\mathrm{kg\,m}^{-3}$ (air)\vspace*{0.2cm}
      \end{tabular}\\

$R$ & specific gas constant & $461.52\,\mathrm{m}^{2}\mathrm{s}^{-2}\mathrm{K}^{-1}$ (water)\vspace*{0.2cm}\\

$\mu_{s.l}$ & shear viscosity (liquid) & $0.890\times10^{-3}\mathrm{Pa\,s}$ (water)\vspace*{0.2cm}\\

$\mu_{s.v}$ & shear viscosity (vapor)\hspace*{0.5cm}
    & \hspace*{-0.2cm}
      \begin{tabular}[c]{l}
      $0.9867\times10^{-5}\,\mathrm{kg\,m}^{-1}\mathrm{s}^{-1}$ (water)\vspace*{-0.1cm}\\
      $1.8374\times10^{-5}\,\mathrm{kg\,m}^{-1}\mathrm{s}^{-1}$ (air)
      \end{tabular}\vspace*{0.2cm}\\

$\mu_{b.v}$ & bulk viscosity (vapor)
    & \hspace*{-0.2cm}
      \begin{tabular}[c]{l}
      $2.7380\times10^{-5}\,\mathrm{kg\,m}^{-1}\mathrm{s}^{-1}$ (water)\vspace*{-0.1cm}\\
      $1.7466\times10^{-5}\,\mathrm{kg\,m}^{-1}\mathrm{s}^{-1}$ (air)
      \end{tabular}\vspace*{0.2cm}\\

$\sigma$ & surface tension & $72.06\,\mathrm{mN}\,\mathrm{m}^{-1}$ (water/air)\vspace*{0.2cm}\\

$K$ & the Korteweg constant & $2.45\times10^{-17}\mathrm{m}^{7}\mathrm{kg}^{-1}\mathrm{s}^{-2}$ (water)

\end{tabular}
\end{ruledtabular}
\end{table*}

Instead, one can treat $l$ and $\bar{h}$ as adjustable parameters and fix
their values by fitting the theoretical results to experimental data (which is
how all other models of contact lines work without exception).

As for $Q(h)$, one can show that, to leading order, it reduces to%
\[
Q(h)=\frac{1}{\mu_{s.l}}\left(  \frac{1}{3}h^{3}+Q_{0}l^{3}\right)  ,
\]
where $Q_{0}$ depends on $G(\rho,T)$ and $\mu_{s}(\rho,T)$. For the van der
Waals fluid under an extra assumption that $\mu_{s}$ is proportional to $\rho
$, the constant $Q_{0}$ happens to be small: $Q_{0}\approx0.098595$.
Furthermore, since Eq. (\ref{5.1}) was derived under the assumption that the
ratio $h/l$ is (logarithmically) large, the second term in the above
expression is small. This claim has been verified by computations: in
particular, the solutions in Figs. \ref{fig7}--\ref{fig9} have turned out to
be indistinguishable from those computed for $Q_{0}=0$ or $Q_{0}%
=2\times0.098595$, with the rest of the parameters being the same. This
suggests that one can simulate Eq. (\ref{5.1}) with simply%
\[
Q(h)=\frac{1}{3\mu_{s.l}}h^{3}.
\]
Note also that $l$, $\bar{h}$, and $Q(h)$ do not appear in the first term on
the right-hand side of Eq. (\ref{5.1}), which describes the Kelvin effect.
This allows one to objectively estimate the characteristic time of the Kelvin
effect, defined as%
\begin{equation}
\tau=\frac{0.14219\left(  \mu_{b.v}+\frac{4}{3}\mu_{s.v}\right)  }{\sigma
}\left(  \dfrac{\rho_{l}}{\rho_{v}}\right)  ^{5/2}\left(  \dfrac{RT}{K\rho
_{l}}\right)  ^{1/2}L^{2}, \label{5.2}%
\end{equation}
where $L$ is the horizontal scale of the liquid film.

To place this estimate in the context of one's everyday experience, $\tau$
will be estimated for the parameters of household mould: it is known to appear
in corners and wall irregularities, and is generally a good example of
Kelvin-effect-induced condensation. Thus, using the parameters of water for
the liquid phase and those of air for the vapor, and letting
$L=0.1\,\mathrm{mm}$ as the smallest visible mould size, one obtains
$\tau\approx11\,\mathrm{h}$. This estimate characterizes how quickly a wet
spot would become visible if the air in one's dwelling is 100\% humid.

One should keep in mind, however, that Eq. (\ref{5.1}) and estimate
(\ref{5.2}) have been derived for a pure fluid -- hence, using them for the
water/air combination is somewhat inconsistent. To obtain a more reliable
estimate, one needs an extension of the present results to multicomponent
fluids, which is currently in progress. One should also take into account the
absorption of the condensate by the wallpaper or plaster, as well as its
consumption by bacteria (which turn the liquid into the actual mould).

Another potential application of the present results is liquid films in steam
turbines, where the temperature can be as high as $400^{\circ}\mathrm{C}$. The
corresponding value of $\rho_{l}/\rho_{v}$ is much smaller than that at room
temperature, so that estimate (\ref{5.2}) predicts that the condensation is
quicker by several orders of magnitude.

\section{Concluding remarks\label{Sec 6}}

This paper examined the evolution of saturated vapor between two intersecting
walls, and its main physical result is condition (\ref{1.1}). If the angle
$\phi$ at which the walls intersects and the contact angle $\theta$ satisfy
this condition, the vapor begins to condensate and a liquid meniscus starts to
grow in the corner. If condition (\ref{1.1}) does \emph{not} hold, there is a
steady (non-growing) solution describing a steady meniscus. Both these results
have been obtained using Eq. (\ref{4.1}) derived in Appendix
\ref{Appendix B.1} under the assumptions that $\theta\ll1$, $\phi\approx\pi$
(hydrophilic walls intersecting at an almost straight angle), and $\rho
_{v}/\rho_{l}\ll1$ (the vapor-to-liquid density ratio is small). For the
regime $\rho_{v}/\rho_{l}\sim1$, a separate asymptotic equation was derived,
Eq. (\ref{B.22}) of Appendix \ref{Appendix B.2}; its solutions have not been
described in this paper, as they are similar to those of Eq. (\ref{4.1}).

The main mathematical result of the present paper are asymptotic equation
(\ref{4.1}) and its dimensional version (\ref{5.1}). They were used to
formally derive condition (\ref{1.1}), but can also be employed for modeling
thin drops with contact lines.

To understand in what way Eq. (\ref{5.1}) differs from the existing
liquid-film models incorporating evaporation (e.g.
\cite{DeeganBakajinDupontHuberEtal00,DunnWilsonDuffyDavidSefiane09,EggersPismen10,ColinetRednikov11,RednikovColinet13,Morris14,StauberWilsonDuffySefiane14,StauberWilsonDuffySefiane15,JanecekDoumencGuerrierNikolayev15,SaxtonWhiteleyVellaOliver16,SaxtonVellaWhiteleyOliver17,BrabcovaMchaleWellsBrown17,RednikovColinet19,WrayDuffyWilson19}%
), note that there are two distinct mechanisms of evaporation of drops:

\begin{enumerate}
\item[(a)] through \emph{diffusion} of vapor in the surrounding air, and

\item[(b)] through \emph{advection} of vapor by the flow due to the variations
of the chemical potential (caused by the curvature of the drop's surface).
\end{enumerate}

All of the existing models are based on mechanism (a), whereas the present
work, on mechanism (b). The latter is the only one acting in \emph{pure}
fluids, where diffusion does not occur.

In \emph{multicomponent} fluids, however, both mechanism should be accounted
for -- but, so far, only (a) has. This shortcoming can be remedied using the
multicomponent DIM -- or perhaps one of the models incorporating kinetic
theory (see \cite{Sazhin17} and references therein).

\acknowledgments{The author is grateful to Demetrios Papageorgiou for a helpful question and to
Mikhail Benilov, for a helpful advice.}

\section*{Author declarations}

\subsection*{Conflict of interest}

The author has no conflicts to disclose.

\subsection*{Data availability}

The data that support the findings of this study are available within the article.

\appendix

\section{Asymptotic description of static menisci\label{Appendix A}}

For simplicity, the asymptotic analysis in both appendices of this paper will
be carried out for two-dimensional (2D) flow. The 3D versions of the equations
derived can be easily deduced afterwards from the requirement of horizontal isotropy.

\subsection{Preliminaries\label{Appendix A.1}}

In what follows, one needs, firstly, an expansion of the thickness $\bar{h}$
of the precursor film, and secondly, the large-distance asymptotics of the
function $\bar{\rho}(z)$ describing a flat interface in an unbounded space.

(1) The Maxwell construction (\ref{2.19})--(\ref{2.20}) and identity
(\ref{2.4}) imply%
\begin{multline}
\rho\left[  G(\rho,T)-G(\rho_{v},T)\right]  -p(\rho,T)+p(\rho_{v},T)\\
=\frac{C^{2}}{2}\left(  \rho-\rho_{l}\right)  ^{2}+\mathcal{O}[\left(
\rho-\rho_{l}\right)  ^{3}]\qquad\text{as}\qquad\rho\rightarrow\rho_{l},
\label{A.1}%
\end{multline}
where $C$ is given by (\ref{3.17}). Expansion (\ref{A.1}) implies that the
integrand in expression (\ref{3.16}) for $\bar{h}$ has a first-order pole at
$\rho=\rho_{l}$; it is located \emph{outside} the integration interval, but
not too far from its upper limit. Thus, (\ref{3.16}) reduces to%
\begin{equation}
\bar{h}=\bar{h}_{0}+\bar{h}_{1}+\mathcal{O}(\varepsilon), \label{A.2}%
\end{equation}
where\begin{widetext}%
\begin{align}
\bar{h}_{0}  & =\frac{\ln\varepsilon^{-1}}{C},\label{A.3}\\
\bar{h}_{1}  & =\int_{\frac{1}{2}\left(  \rho_{l}+\rho_{v}\right)  }^{\rho
_{l}}\left\{  \frac{2^{-1/2}}{\sqrt{\rho\left[  G(\rho,T)-G(\rho
_{v},T)\right]  -p(\rho,T)+p(\rho_{v},T)}}-\frac{1}{C\left(  \rho_{l}%
-\rho\right)  }\right\}  \mathrm{d}\rho+\frac{1}{C}\ln\frac{\rho_{l}-\rho_{v}%
}{2}.\label{A.4}%
\end{align}
\end{widetext}

(2) It follows from the exact solution (\ref{3.14}) that%
\[
\bar{\rho}(z)\sim\rho_{l}-\operatorname{e}^{C\left(  z+\bar{h}_{1}\right)
}\qquad\text{as}\qquad z\rightarrow-\infty,
\]
where $\bar{h}_{1}$ is given by (\ref{A.4}). Then, using equalities
(\ref{A.2})--(\ref{A.3}), one can rewrite the above estimate in terms of the
full thickness of the precursor film,%
\begin{equation}
\bar{\rho}(z)\sim\rho_{l}-\varepsilon\operatorname{e}^{C\left(  z+\bar
{h}\right)  }\qquad\text{as}\qquad z\rightarrow-\infty. \label{A.5}%
\end{equation}
This expansion holds as long as its second term is smaller than the first one
-- i.e., for moderately (logarithmically) large distances, $-z\gtrsim\bar{h}$.

\subsection{Derivation of Eq. (\ref{3.24})\label{Appendix A.2}}

The solution of Eq. (\ref{3.6}) will be sought in the form%
\begin{equation}
\rho(x,z)=\bar{\rho}(z-H-h)+\varepsilon^{2}\rho^{(2)}+\cdots, \label{A.6}%
\end{equation}
where $\bar{\rho}(z)$ describes a flat interface in an unbounded space and
satisfies boundary-value problem (\ref{3.9})--(\ref{3.12}). Physically,
solution (\ref{A.6}) describes a slightly curved interface located at
$z=H(x)+h(x)$ (i.e., at a height $h(x)$ above the substrate), and a small
correction. In what follows, the two-term expansion (\ref{A.6}) plays an
important role, for both static and evolving menisci.

Substituting (\ref{A.6}) into Eq. (\ref{3.6}) and boundary condition
(\ref{3.8}) (boundary condition (\ref{3.7}) will be discussed later), one
obtains%
\begin{equation}
\frac{\partial^{2}\rho^{(2)}}{\partial z^{2}}-\left(  \frac{\partial
G}{\partial\rho}\right)  _{\rho=\bar{\rho}}\rho^{(2)}=\mathcal{R}, \label{A.7}%
\end{equation}%
\begin{equation}
\rho^{(2)}\rightarrow0\qquad\text{as}\qquad z\rightarrow\infty, \label{A.8}%
\end{equation}
where%
\begin{equation}
\mathcal{R}=-\frac{\partial^{2}\bar{\rho}}{\partial x^{2}}, \label{A.9}%
\end{equation}
and it is implied here (and in the rest of the paper, unless stated otherwise)
that $\bar{\rho}$ depends on $z-H-h$, not just $z$.

(\ref{A.7}) is a linear nonhomogeneous second-order ordinary differential
equation, and it can be readily verified that its homogeneous version is
satisfied by $\rho^{(2)}=\partial\bar{\rho}/\partial z$. Thus, its general
solution is easy to find: imposing boundary condition (\ref{A.8}), one
obtains, after straightforward algebra,%
\begin{multline}
\rho^{(2)}=\frac{\partial\bar{\rho}}{\partial z}\left[  \int_{z}^{H+h}\left(
\frac{\partial\bar{\rho}^{\prime}}{\partial z}\right)  ^{-2}F(x,z^{\prime
})\,\mathrm{d}z^{\prime}\right. \\
\left.  -h^{(2)}%
(x)\vphantom{\int_{z}^{H+h}\left( \frac{\partial\bar{\rho}^{\prime}}{\partial z}\right) ^{-2}}\right]
, \label{A.10}%
\end{multline}
where $\bar{\rho}^{\prime}=\bar{\rho}(z^{\prime}-H-h)$,%
\begin{equation}
F(x,z)=\int_{z}^{\infty}\frac{\partial\bar{\rho}^{\prime}}{\partial z^{\prime
}}\mathcal{R}(x,z^{\prime})\,\mathrm{d}z^{\prime}, \label{A.11}%
\end{equation}
and the undetermined function $h^{(2)}(x)$ is, mathematically, a constant of
integration. Physically, $h^{(2)}$ corresponds to shifting the interface along
the $z$ axis by a distance of $\varepsilon^{2}h^{(2)}$. In principle, it can
be eliminated by replacing in expansion (\ref{A.6}) the leading-order solution
$\bar{\rho}(z-H-h)$ with $\bar{\rho}(z-H-h-\varepsilon^{2}h^{(2)})$.

Expansion (\ref{A.6}) is valid if its second term $\varepsilon^{2}\rho^{(2)}$
is much smaller than the first term $\bar{\rho}$. This requirement clearly
holds near the interface, where $z-H-h=\mathcal{O}(1)$ -- hence, $\rho^{(2)}$
does not involve any small or large parameters -- hence, $\rho^{(2)}%
=\mathcal{O}(1)$. Furthermore, as shown below, the condition $\varepsilon
^{2}\rho^{(2)}\ll\bar{\rho}$ holds near the substrate as well (even though
$z-H-h$ can be large there, and so can $\rho^{(2)}$).

Thus, since expansion (\ref{A.6}) is uniformly applicable, there is no need to
introduce a near-substrate boundary layer. Substituting expression
(\ref{A.10}) into (\ref{A.6}), then substituting the latter into boundary
condition (\ref{3.7}), one obtains\begin{widetext}%
\begin{equation}
\left(  \bar{\rho}\right)  _{z=H}+\varepsilon^{2}\left(  \frac{\partial
\bar{\rho}}{\partial z}\right)  _{z=H}\left[  \int_{H}^{H+h}\left(
\frac{\partial\bar{\rho}}{\partial z}\right)  ^{-2}F(x,z)\,\mathrm{d}%
z-h^{(2)}\right]  =\rho_{l}-\varepsilon.\label{A.12}%
\end{equation}
This is, essentially, the desired equation for $h(x)$. To reduce it to
Eq. (\ref{3.24}), one should assume that $h$ is logarithmically large.
Physically, such an assumption is not restrictive, as liquid menisci are
indeed thicker than the precursor film (describing \emph{dry} substrate), and
the thickness $\bar{h}$ of the latter \emph{is} logarithmically large due to
estimates (\ref{A.2})--(\ref{A.4}).
For large $h$, the main contribution to the integral in (\ref{A.12}) comes
from the region adjacent to its lower limit, where $\partial\bar{\rho
}/\partial z$ is small. Thus, one can use (\ref{A.5}) to obtain%
\begin{equation}
\int_{H}^{H+h}\left(  \frac{\partial\bar{\rho}}{\partial z}\right)
^{-2}F(x,z)\,\mathrm{d}z=\frac{\varepsilon^{-2}}{2C^{3}}\operatorname{e}%
^{-2C(h-\bar{h})}F(x,H)+\mathcal{O}(\ln\varepsilon^{-1}).\label{A.13}%
\end{equation}
Given this estimate and (\ref{A.5}), the second term on the left-hand side of
Eq. (\ref{A.12}) is $\mathcal{O}(\varepsilon)$ -- hence, it is much
smaller than the first term. This justifies the use of expansion (\ref{A.6})
near the substrate.
Substituting estimate (\ref{A.13}) into Eq. (\ref{A.12}) and using the
large-distance asymptotics (\ref{A.5}) of $\bar{\rho}$ to simplify the rest of
(\ref{A.12}), one can reduce it to leading order to%
\begin{equation}
F(x,H)=2C^{2}\left[  1-\operatorname{e}^{-C(h-\bar{h})}\right]
\operatorname{e}^{-C(h-\bar{h})}\mathbf{.} \label{A.14}%
\end{equation}
Note that the undetermined function $h^{(2)}(x)$ does not appear in this
(leading-order) equation; it can only be determined in the next order of the
perturbation expansion.
To close Eq. (\ref{A.14}), it remains to express $F$ in terms of $h$.
Backtracking through equalities (\ref{A.11}) and (\ref{A.9}) (thus, relating
$F$ to $\mathcal{R}$ to $\bar{\rho}$), one obtains%
\begin{equation}
\int_{H}^{\infty}\frac{\partial\bar{\rho}}{\partial z}\left\{  \frac
{\mathrm{d}^{2}(H+h)}{\mathrm{d}z^{2}}\frac{\partial\bar{\rho}}{\partial
z}-\left[  \frac{\mathrm{d}(H+h)}{\mathrm{d}z}\right]  ^{2}\frac{\partial
^{2}\bar{\rho}}{\partial z^{2}}\right\}  \mathrm{d}z=2C^{2}\left[
1-\operatorname{e}^{-C(h-\bar{h})}\right]  \operatorname{e}^{-C(h-\bar{h}%
)}.\label{A.15}%
\end{equation}
Observe that, since $h$ is large, it follows from the large-distance
asymptotics (\ref{A.5}) that%
\[
\frac{\partial\bar{\rho}(z-H-h)}{\partial z}=\mathcal{O}(\varepsilon
),\qquad\frac{\partial^{2}\bar{\rho}(z-H-h)}{\partial z^{2}}=\mathcal{O}%
(\varepsilon)\qquad\text{if}\qquad z<H.
\]
\end{widetext}As a result, the lower limit of the integral in (\ref{A.15}) can
be moved from $H$ to $-\infty$ without introducing a leading-order error.
After that, the second term on the left-hand side of Eq. (\ref{A.15})
vanishes, and (\ref{A.15}) turns into Eq. (\ref{3.24}) as required.

\section{Asymptotic equations for evolving menisci\label{Appendix B}}

As mentioned in the main body of the paper, there are two asymptotic regimes
in this problem, depending on the parameter $\rho_{v}/\rho_{l}$. The common
part of their analyses will be presented first, with the regime-specific parts
to follow.

To nondimensionalize the governing equations, assume that the shear and bulk
viscosities are of the same order, $\mu_{s}\sim\mu_{b}$, and introduce a scale
$\mu$ representing them both. As shown by \cite{Benilov20d} for a flat
substrate, the scale for the horizontal velocity $u$ is determined by the
balance of the viscous and Korteweg stresses, so that%
\begin{equation}
U=\frac{\varepsilon^{3}Pl}{\mu}, \label{B.1}%
\end{equation}
where $\varrho$, $P$, and $l$ have been defined in the beginning of
\S \ref{Sec 3}. The scale for the vertical velocity $w$ is regime-specific and
will be chosen later -- as is, and will be, the time scale.

In addition to the nondimensional variables defined by (\ref{3.3}%
)--(\ref{3.5}), introduce%
\begin{equation}
u_{nd}=\frac{u}{U}, \label{B.2}%
\end{equation}%
\begin{equation}
c_{nd}=\frac{c}{R},\qquad B_{nd}=\frac{B}{P},\qquad s_{nd}=\frac{s}{R},
\label{B.2a}%
\end{equation}%
\begin{equation}
\left(  \mu_{s}\right)  _{nd}=\frac{\mu_{s}}{\mu},\qquad\left(  \mu
_{b}\right)  _{nd}=\frac{\mu_{b}}{\mu},\qquad\kappa_{nd}=\frac{\kappa
}{\varkappa}, \label{B.3}%
\end{equation}
where $\varkappa$ is a characteristic scale of the thermal conductivity, and
the specific gas constant $R$ is that of the specific heat capacity $c$.

\subsection{Regime 1: $\rho_{v}\sim\rho_{l}$\label{Appendix B.1}}

Let the nondimensional time and vertical velocity be%
\begin{equation}
t_{nd}=\frac{t}{l/\left(  \varepsilon^{-1}U\right)  },\qquad w_{nd}=\frac
{w}{\varepsilon^{-1}U}. \label{B.4}%
\end{equation}
Substituting (\ref{3.3})--(\ref{3.5}) and (\ref{B.2})--(\ref{B.4}) into the 2D
version of boundary-value problem (\ref{2.8})--(\ref{2.18}), and omitting the
subscript $_{nd}$, one obtains%
\begin{equation}
\frac{\partial\rho}{\partial t}+\varepsilon^{2}\frac{\partial\left(  \rho
u\right)  }{\partial x}+\frac{\partial\left(  \rho w\right)  }{\partial z}=0,
\label{B.5}%
\end{equation}
\begin{widetext}%
\begin{multline}
\alpha\varepsilon^{4}\left(  \frac{\partial u}{\partial t}+\varepsilon
^{2}u\frac{\partial u}{\partial x}+w\frac{\partial u}{\partial z}\right)
+s\frac{\partial T}{\partial x}+\frac{\partial}{\partial x}\left(
G-\varepsilon^{2}\frac{\partial^{2}\rho}{\partial x^{2}}-\frac{\partial
^{2}\rho}{\partial z^{2}}\right)  \\
=\frac{\varepsilon^{2}}{\rho}\left\{  \frac{\partial}{\partial x}\left[
2\varepsilon^{2}\mu_{s}\frac{\partial u}{\partial x}+\left(  \mu_{b}%
-\frac{2\mu_{s}}{3}\right)  \left(  \varepsilon^{2}\frac{\partial u}{\partial
x}+\frac{\partial w}{\partial z}\right)  \right]  +\frac{\partial}{\partial
z}\left[  \mu_{s}\left(  \frac{\partial u}{\partial z}+\frac{\partial
w}{\partial x}\right)  \right]  \right\}  ,\label{B.6}%
\end{multline}%
\begin{multline}
\alpha\varepsilon^{4}\left(  \frac{\partial w}{\partial t}+\varepsilon
^{2}u\frac{\partial w}{\partial x}+w\frac{\partial w}{\partial z}\right)
+s\frac{\partial T}{\partial z}+\frac{\partial}{\partial z}\left(
G-\varepsilon^{2}\frac{\partial^{2}\rho}{\partial x^{2}}-\frac{\partial
^{2}\rho}{\partial z^{2}}\right)  \\
=\frac{\varepsilon^{4}}{\rho}\frac{\partial}{\partial x}\left[  \mu_{s}\left(
\frac{\partial u}{\partial z}+\frac{\partial w}{\partial x}\right)  \right]
+\frac{\varepsilon^{2}}{\rho}\frac{\partial}{\partial z}\left[  2\mu_{s}%
\frac{\partial w}{\partial z}+\left(  \mu_{b}-\frac{2\mu_{s}}{3}\right)
\left(  \varepsilon^{2}\frac{\partial u}{\partial x}+\frac{\partial
w}{\partial z}\right)  \right]  ,\label{B.7}%
\end{multline}%
\begin{multline}
\alpha\gamma\rho c\left(  \frac{\partial T}{\partial t}+\varepsilon^{2}%
u\frac{\partial T}{\partial x}+w\frac{\partial T}{\partial z}\right)  +\beta
B\left(  \varepsilon^{2}\frac{\partial u}{\partial x}+\frac{\partial
w}{\partial z}\right)  \\
=\beta\varepsilon^{2}\left\{  \mu_{s}\left[  2\varepsilon^{4}\left(
\frac{\partial u}{\partial x}\right)  ^{2}+\varepsilon^{2}\left(
\frac{\partial u}{\partial z}+\frac{\partial w}{\partial x}\right)
^{2}+2\left(  \frac{\partial w}{\partial z}\right)  ^{2}\right]  +\left(
\mu_{b}-\frac{2\mu_{s}}{3}\right)  \left(  \varepsilon^{2}\frac{\partial
u}{\partial x}+\frac{\partial w}{\partial z}\right)  ^{2}\right\}  \\
+\frac{\partial}{\partial x}\left(  \kappa\frac{\partial T}{\partial
x}\right)  +\dfrac{1}{\varepsilon^{2}}\frac{\partial}{\partial z}\left(
\kappa\frac{\partial T}{\partial z}\right)  ,\label{B.8}%
\end{multline}
\end{widetext}%
\begin{equation}
u=0,\qquad w=0\qquad\text{at}\qquad z=H, \label{B.9}%
\end{equation}%
\begin{equation}
\frac{\partial u}{\partial z}\rightarrow0,\qquad\frac{\partial w}{\partial
z}\rightarrow0\qquad\text{as}\qquad z\rightarrow\infty, \label{B.10}%
\end{equation}%
\begin{equation}
\rho=\rho_{l}-\varepsilon,\qquad T=T_{0}\qquad\text{at}\qquad z=H,
\label{B.11}%
\end{equation}%
\begin{equation}
\rho\rightarrow\rho_{v},\qquad\frac{\partial T}{\partial z}\rightarrow
0\qquad\text{as}\qquad z\rightarrow\infty, \label{B.12}%
\end{equation}
where%
\[
\alpha=\frac{K\varrho^{3}}{\mu^{2}},\qquad\beta=\dfrac{P\varrho^{2}K}%
{\mu\varkappa\left(  T_{0}\right)  _{d}},
\]%
\[
\gamma=\frac{R\mu}{\varkappa},\qquad T_{0}=\frac{\varrho R\left(
T_{0}\right)  _{d}}{P},
\]
and $\left(  T_{0}\right)  _{d}$ is the dimensional temperature of the substrate.

The positions where $\alpha$ appears in Eqs. (\ref{B.6})--(\ref{B.7}) suggest
that it represents the Reynolds number, $\beta$ is the \textquotedblleft
isothermality parameter\textquotedblright\ introduced by \cite{Benilov20b},
and $\gamma$ is the Prandtl number. For generality, these parameters are
assumed to be order one (as shown by \cite{Benilov20b}, they are typically
either that or small).

Observe that Eq. (\ref{B.8}) involves a term proportional to $1/\varepsilon
^{2}$, which cancels only if%
\begin{equation}
T=T_{0}+\varepsilon^{2}\tilde{T}, \label{B.13}%
\end{equation}
i.e., the temperature variations are small. This does not mean, however, that
their effect on the film dynamics is negligible. To make it such, one should
also assume the isothermality parameter $\beta$ to be also small
\cite{Benilov20d}.

Substituting (\ref{B.13}) into Eq. (\ref{B.7}) and simplifying the notation by
changing\thinspace$T_{0}\rightarrow T$, one obtains\begin{widetext}%
\begin{equation}
\frac{\partial}{\partial z}\left(  \frac{\partial^{2}\rho}{\partial z^{2}%
}-G\right)  =\varepsilon^{2}\left[  \frac{\partial}{\partial z}\left(
-\frac{\partial^{2}\rho}{\partial x^{2}}+\frac{\partial G}{\partial T}%
\tilde{T}\right)  +s(T,\rho)\frac{\partial\tilde{T}}{\partial z}-\frac{1}%
{\rho}\frac{\partial}{\partial z}\left(  \lambda\frac{\partial w}{\partial
z}\right)  \right]  +\mathcal{O}(\varepsilon^{4}),\label{B.14}%
\end{equation}
where the effective viscosity $\lambda(\rho,T)$ is given by%
\begin{equation}
\lambda(\rho,T)=\mu_{b}(\rho,T)+\frac{4}{3}\mu_{s}(\rho,T).\label{B.15}%
\end{equation}
For evolving menisci, the film thickness $h$ depends on $x$ and $t$ (not just
on $x$ as in the static case). Keeping this in mind whilst substituting
(\ref{A.6}) into (\ref{B.14}), one obtains%
\begin{equation}
\frac{\partial^{2}\rho^{(2)}}{\partial z^{2}}-\left(  \frac{\partial
G}{\partial\rho}\right)  _{\rho=\bar{\rho}}=\varepsilon^{2}\mathcal{R}%
_{1},\label{B.16}%
\end{equation}
where%
\begin{equation}
\mathcal{R}_{1}=-\frac{\partial^{2}\bar{\rho}}{\partial x^{2}}+\frac{\partial
G(\bar{\rho},T)}{\partial T}\tilde{T}-\frac{\partial G(\rho_{v},T)}{\partial
T}\left(  \tilde{T}\right)  _{z\rightarrow\infty}-\int_{z}^{\infty}\left\{
s(\bar{\rho}^{\prime},T)\frac{\partial\tilde{T}^{\prime}}{\partial z^{\prime}%
}-\frac{1}{\bar{\rho}^{\prime}}\frac{\partial}{\partial z^{\prime}}\left[
\lambda(\bar{\rho}^{\prime},T)\frac{\partial w^{\prime}}{\partial z^{\prime}%
}\right]  \right\}  \mathrm{d}z^{\prime},\label{B.17}%
\end{equation}and $\tilde{T}^{\prime}=\tilde{T}(x,z^{\prime},t)$, $w^{\prime
}=w(x,z^{\prime},t)$, etc.
Evidently, the left-hand side of Eq. (\ref{B.16}) coincides with that of
its static counterpart (\ref{A.7}), and the boundary conditions for the two
equations also coincide. Thus, the asymptotic equation for evolving menisci
can be obtained by simply replacing $\mathcal{R}$ with $\mathcal{R}_{1}$ in
the static equations (\ref{A.11}) and (\ref{A.14}), which yield%
\begin{equation}
\int_{H}^{\infty}\frac{\partial\bar{\rho}}{\partial z}\mathcal{R}%
_{1}(x,z)\,\mathrm{d}z=2C^{2}\left[  1-\operatorname{e}^{-C(h-\bar{h}%
)}\right]  \operatorname{e}^{-C(h-\bar{h})}\mathbf{.}\label{B.18}%
\end{equation}
Next, substitute expression (\ref{B.17}) for $\mathcal{R}_{1}$ into the above
equality, and then eliminate the integration with respect to $z^{\prime}$ by
integrating by parts the term involving curly brackets. Recall also that $h$
is large (as assumed in Appendix \ref{Appendix A}), which implies%
\begin{equation}
\left[  \bar{\rho}(z-H-h)\right]  _{z=H}=\rho_{l}+\mathcal{O}(\varepsilon
).\label{B.19}%
\end{equation}
Thus, to leading order, one can rearrange (\ref{B.18}) into\begin{multline}
-\int_{H}^{\infty}\frac{\partial\bar{\rho}}{\partial z}\frac{\partial^{2}%
\bar{\rho}}{\partial x^{2}}\mathrm{d}z+\int_{H}^{\infty}\left[  \frac
{\partial\bar{\rho}}{\partial z}\frac{\partial G(\bar{\rho},T)}{\partial
T}\tilde{T}-\bar{\rho}s(\bar{\rho},T)\frac{\partial\tilde{T}}{\partial
z}\right]  \mathrm{d}z\\
+\rho_{l}\int_{H}^{\infty}s(\bar{\rho},T)\frac{\partial\tilde{T}}{\partial
z}\mathrm{d}z-\left(  \rho_{v}-\rho_{l}\right)  \frac{\partial G(\rho_{v}%
,T)}{\partial T}\left(  \tilde{T}\right)  _{z\rightarrow\infty}-\rho_{l}%
\int_{H}^{\infty}\frac{\lambda(\bar{\rho},T)}{\bar{\rho}^{2}}\frac
{\partial\bar{\rho}}{\partial z}\frac{\partial w}{\partial z}\mathrm{d}z\\
=2C^{2}\left[  1-\operatorname{e}^{-C(h-\bar{h})}\right]  \operatorname{e}%
^{-C(h-\bar{h})}.\label{B.20}%
\end{multline}To obtain a closed equation for $h$, the unknowns $\tilde{T}$
and $w$ should be expressed in terms of $\bar{\rho}(z-H-h)$ -- which is not
difficult, as it needs to be done to leading order only. Retaining, thus, the
leading-order terms in Eqs. (\ref{B.5}), (\ref{B.8}), and (\ref{B.9}%
)--(\ref{B.12}), and changing $T_{0}\rightarrow T$, one obtains%
\[
\frac{\partial\bar{\rho}}{\partial t}+\frac{\partial\left(  \bar{\rho
}w\right)  }{\partial z}=0,
\]%
\[
-\beta B(\bar{\rho},T)\frac{\partial w}{\partial z}+\frac{\partial}{\partial
z}\left[  \kappa(\bar{\rho},T)\frac{\partial\tilde{T}}{\partial z}\right]  =0,
\]%
\[
w=0,\qquad\tilde{T}=0\qquad\text{at}\qquad z=H,
\]%
\[
\frac{\partial w}{\partial z}\rightarrow0,\qquad\frac{\partial\tilde{T}%
}{\partial z}\rightarrow0\qquad\text{as}\qquad z\rightarrow\infty.
\]
Keeping in mind estimate (\ref{B.19}) and recalling definition (\ref{2.7}) of
$B(\rho,T)$, one can deduce that, to leading order,\begin{equation}
w=\frac{\partial h}{\partial t}\frac{\bar{\rho}-\rho_{l}}{\bar{\rho}}%
,\qquad\tilde{T}=-\beta T\rho_{l}\frac{\partial h}{\partial t}\int_{H}%
^{z}\frac{s(\bar{\rho}^{\prime},T)-s(\rho_{v},T)}{\kappa(\bar{\rho}^{\prime
},T)}\mathrm{d}z^{\prime}.\label{B.21}%
\end{equation}
Substituting these expressions into Eq. (\ref{B.20}), one obtains, after
straightforward algebra,%
\begin{equation}
\left(  A+\beta D\right)  \frac{\partial h}{\partial t}=\sigma\frac
{\partial^{2}(H+h)}{\partial x^{2}}-2C^{2}\left[  1-\operatorname{e}%
^{-C(h-\bar{h})}\right]  \operatorname{e}^{-C(h-\bar{h})},\label{B.22}%
\end{equation}
where%
\begin{equation}
\sigma=\int_{H}^{\infty}\left(  \frac{\partial\bar{\rho}}{\partial z}\right)
^{2}\mathrm{d}z,\qquad A=\rho_{l}^{2}\int_{H}^{\infty}\frac{\lambda(\bar{\rho
},T)}{\bar{\rho}^{4}}\left(  \frac{\partial\bar{\rho}}{\partial z}\right)
^{2}\mathrm{d}z,\label{B.23}%
\end{equation}%
\begin{equation}
D=T\rho_{l}^{2}\int_{H}^{\infty}\left[  s(\bar{\rho},T)-s(\rho_{v}%
,T)-\frac{\rho_{l}-\rho_{v}}{\rho_{l}}\left(  \rho\frac{\partial s}%
{\partial\rho}\right)  _{\rho=\rho_{v}}\right]  \frac{s(\bar{\rho}%
,T)-s(\rho_{v},T)}{\kappa(\bar{\rho},T)}\mathrm{d}z.\label{B.24}%
\end{equation}
(\ref{B.22}) is the desired asymptotic equation describing menisci with
order-one vapor-to-liquid density ratio, but its coefficients can be
simplified further, similar to how it was done in Appendix \ref{Appendix A}.
Moving the lower limit of integration in (\ref{B.23}) from $H$ to $-\infty$,
one can reduce $\sigma$ to its standard form, (\ref{3.25}), and $A$, to%
\begin{equation}
A=\rho_{l}^{2}\int_{-\infty}^{\infty}\frac{\lambda(\bar{\rho},T)}{\bar{\rho
}^{4}}\left(  \frac{\partial\bar{\rho}}{\partial z}\right)  ^{2}%
\mathrm{d}z.\label{B.25}%
\end{equation}
The integrand in (\ref{B.24}), in turn, tends to a constant as $z\rightarrow
-\infty$, so the lower limit cannot be moved to $-\infty$. One can still
simplify (\ref{B.24}) by integrating it by parts and \emph{then} moving the
limit to $-\infty$. Eventually, (\ref{B.24}) becomes%
\[
D=D_{1}h-D_{2},
\]
where%
\[
D_{1}=T\rho_{l}^{2}\left[  s(\rho_{l},T)-s(\rho_{v},T)-\left(  1-\frac
{\rho_{v}}{\rho_{l}}\right)  \left(  \rho\frac{\partial s}{\partial\rho
}\right)  _{\rho=\rho_{v}}\right]  \frac{s(\rho_{l},T)-s(\rho_{v},T)}%
{\kappa(\rho_{l},T)},
\]%
\[
D_{2}=T\rho_{l}^{2}\int_{-\infty}^{\infty}z\frac{\partial}{\partial z}\left\{
\left[  s(\bar{\rho},T)-s(\rho_{v},T)-\frac{\rho_{l}-\rho_{v}}{\rho_{l}%
}\left(  \rho\frac{\partial s}{\partial\rho}\right)  _{\rho=\rho_{v}}\right]
\frac{s(\bar{\rho},T)-s(\rho_{v},T)}{\kappa(\bar{\rho},T)}\right\}
\mathrm{d}z.
\]
\end{widetext}

\subsection{Regime 2: $\rho_{v}\ll\rho_{l}$\label{Appendix B.2}}

Let $\rho_{v}$ be small. Then, according to (\ref{B.21}), the vertical
velocity $w$ in the vapor phase is large. This makes sense physically: a large
density difference between vapor and liquid implies a faster evaporative flow.
Mathematically though, the growth of $w$ makes the scaling inconsistent,
suggesting that a boundary layer exists between the asymptotic regions
describing liquid and vapor.

Thus, three asymptotic regions are expected to arise in the problem: the
\underline{liquid region} where $\rho\sim\rho_{l}$, the \underline{boundary
layer} where $\rho\sim\rho_{v}$, and the \underline{vapor region} where
$\rho\approx\rho_{v}$. The last one is trivial and has no impact on the global
dynamics -- hence, will not be discussed.

\subsubsection{The liquid region\label{Appendix B.2.1}}

The nondimensional time and vertical velocity in this region are defined by%
\begin{equation}
t_{nd}=\frac{t}{l/\left(  \varepsilon U\right)  },\qquad w_{nd}=\frac
{w}{\varepsilon U}, \label{B.26}%
\end{equation}
where $U$ is given by (\ref{B.1}). Substituting (\ref{3.3})--(\ref{3.5}),
(\ref{B.2})--(\ref{B.3}), and (\ref{B.26}) into the 2D version of the
governing set (\ref{2.8})--(\ref{2.12}), one obtains (the subscript $_{nd}$
omitted),%
\begin{equation}
\frac{\partial\rho}{\partial t}+\frac{\partial\left(  \rho u\right)
}{\partial x}+\frac{\partial\left(  \rho w\right)  }{\partial z}=0,
\label{B.27}%
\end{equation}
\begin{widetext}%
\begin{multline}
\alpha\varepsilon^{6}\left(  \frac{\partial u}{\partial t}+u\frac{\partial
u}{\partial x}+w\frac{\partial u}{\partial z}\right)  +s\frac{\partial
T}{\partial x}+\frac{\partial}{\partial x}\left(  G-\varepsilon^{2}%
\frac{\partial^{2}\rho}{\partial x^{2}}-\frac{\partial^{2}\rho}{\partial
z^{2}}\right)  \\
=\frac{\varepsilon^{4}}{\rho}\frac{\partial}{\partial x}\left[  2\mu_{s}%
\frac{\partial u}{\partial x}+\left(  \mu_{b}-\frac{2\mu_{s}}{3}\right)
\left(  \frac{\partial u}{\partial x}+\frac{\partial w}{\partial z}\right)
\right]  +\frac{\varepsilon^{2}}{\rho}\frac{\partial}{\partial z}\left[
\mu_{s}\left(  \frac{\partial u}{\partial z}+\varepsilon^{2}\frac{\partial
w}{\partial x}\right)  \right]  ,\label{B.28}%
\end{multline}%
\begin{multline}
\alpha\varepsilon^{8}\left(  \frac{\partial w}{\partial t}+u\frac{\partial
w}{\partial x}+w\frac{\partial w}{\partial z}\right)  +s\frac{\partial
T}{\partial z}+\frac{\partial}{\partial z}\left(  G-\varepsilon^{2}%
\frac{\partial^{2}\rho}{\partial x^{2}}-\frac{\partial^{2}\rho}{\partial
z^{2}}\right)  \\
=\frac{\varepsilon^{4}}{\rho}\left\{  \frac{\partial}{\partial x}\left[
\mu_{s}\left(  \frac{\partial u}{\partial z}+\varepsilon^{2}\frac{\partial
w}{\partial x}\right)  \right]  +\frac{\partial}{\partial z}\left[  2\mu
_{s}\frac{\partial w}{\partial z}+\left(  \mu_{b}-\frac{2\mu_{s}}{3}\right)
\left(  \frac{\partial u}{\partial x}+\frac{\partial w}{\partial z}\right)
\right]  \right\}  ,\label{B.29}%
\end{multline}%
\begin{multline}
\alpha\gamma\rho c\left(  \frac{\partial T}{\partial t}+u\frac{\partial
T}{\partial x}+w\frac{\partial T}{\partial z}\right)  +\beta B\left(
\frac{\partial u}{\partial x}+\frac{\partial w}{\partial z}\right)  \\
=\beta\varepsilon^{2}\left\{  \mu_{s}\left[  2\varepsilon^{2}\left(
\frac{\partial u}{\partial x}\right)  ^{2}+\left(  \frac{\partial u}{\partial
z}+\varepsilon^{2}\frac{\partial w}{\partial x}\right)  ^{2}+2\varepsilon
^{2}\left(  \frac{\partial w}{\partial z}\right)  ^{2}\right]  +\varepsilon
^{2}\left(  \mu_{b}-\frac{2\mu_{s}}{3}\right)  \left(  \frac{\partial
u}{\partial x}+\frac{\partial w}{\partial z}\right)  ^{2}\right\}  \\
+\frac{1}{\varepsilon^{2}}\frac{\partial}{\partial x}\left(  \kappa
\frac{\partial T}{\partial x}\right)  +\frac{1}{\varepsilon^{4}}\frac
{\partial}{\partial z}\left(  \kappa\frac{\partial T}{\partial z}\right)
.\label{B.30}%
\end{multline}
The boundary conditions for this set coincide with their Regime-1 counterparts
(\ref{B.9})--(\ref{B.12}).
The temperature equation (\ref{B.30}) suggests that
\begin{equation}
T=T_{0}+\mathcal{O}(\varepsilon^{4}), \label{B.31}%
\end{equation}
with the implication that the temperature variations are too small to affect
the leading-order dynamics [compare (\ref{B.31}) to its Regime-1 counterpart
(\ref{B.13})]. Thus, the temperature equation can be simply omitted, and in
the rest of the governing set, one can let $T=\operatorname{const}$.
Given the quasi-isothermality condition (\ref{B.31}), it follows from equation
(\ref{B.29}) that%
\begin{equation}
G(\rho,T)-G(\rho_{v},T)-\varepsilon^{2}\frac{\partial^{2}\rho}{\partial x^{2}%
}-\frac{\partial^{2}\rho}{\partial z^{2}}=\varepsilon^{2}G_{0}+\mathcal{O}%
(\varepsilon^{4}),\label{B.32}%
\end{equation}
where $G_{0}(x,t)$ is an undetermined function. To relate it to $h(x,t)$, one
should use the two-term expansion (\ref{A.6}), after which (\ref{B.32}) yields%
\begin{equation}
\frac{\partial^{2}\rho^{(2)}}{\partial z^{2}}-\left(  \frac{\partial
G}{\partial\rho}\right)  _{\rho=\bar{\rho}}=-\frac{\partial^{2}\bar{\rho}%
}{\partial x^{2}}-G_{0}.\label{B.33}%
\end{equation}
Following the same reasoning as that in Regime 1, but keeping in mind that,
this time, $\rho_{v}\ll1$, one can deduce from the boundary condition for
$\rho$ at the substrate that%
\begin{equation}
G_{0}=-\frac{1}{\rho_{l}}\left\{  \sigma\frac{\partial^{2}(H+h)}{\partial
x^{2}}-2C^{2}\left[  1-\operatorname{e}^{-C(h-\bar{h})}\right]
\operatorname{e}^{-C(h-\bar{h})}\right\}  \mathbf{.}\label{B.34}%
\end{equation}
\end{widetext}Next, substitute (\ref{B.31})--(\ref{B.32}) into Eq.
(\ref{B.28}) for $u$. Keeping in mind that $u$ should satisfy the no-slip
condition at the substrate and the no-viscous-stress condition at infinity,
one obtains%
\begin{equation}
u=-\frac{\partial G_{0}}{\partial x}\int_{H}^{z}\frac{\hat{\rho}^{\prime}}%
{\mu_{s}(\bar{\rho}^{\prime},T)}\mathrm{d}z^{\prime}+\mathcal{O}%
(\varepsilon^{2}), \label{B.35}%
\end{equation}
where $\hat{\rho}$ is related to $\bar{\rho}$ by equality (\ref{4.4}).
Finally, substitute (\ref{B.35}) into Eq. (\ref{B.27}) and solve the latter
for $w$, subject to the no-through-flow requirement at the substrate,%
\begin{multline}
w=\frac{\partial h}{\partial t}\left(  1-\frac{\rho_{l}}{\bar{\rho}}\right) \\
+\frac{1}{\bar{\rho}}\frac{\partial}{\partial x}\left[  \frac{\partial G_{0}%
}{\partial x}\int_{H}^{z}\frac{\left(  \hat{\rho}^{\prime}-\hat{\rho}\right)
\hat{\rho}^{\prime}}{\mu_{s}(\bar{\rho}^{\prime},T)}\mathrm{d}z^{\prime
}\right]  +\mathcal{O}(\varepsilon^{2}). \label{B.36}%
\end{multline}
If the $\bar{\rho}$ is small, this expression is evidently large --
suggesting, as expected, the existence of a boundary layer.

\subsubsection{The boundary layer\label{Appendix B.2.2}}

To obtain the solution in the boundary layer, the liquid-region equations
(\ref{B.27})--(\ref{B.30}) need to be rescaled. The parameters of the new
variables have to be first guessed, then verified through matching to the
liquid region.

The boundary layer will be described by the following inner variables:%
\[
z_{i}=\frac{z-Z(x,t)}{\rho_{v}^{1/2}},\qquad\rho_{i}=\frac{\rho}{\rho_{v}},
\]
where $Z(x,t)$ is the height of the boundary layer. The horizontal coordinate
$x$ and time $t$ do not need to be rescaled (their scales in the boundary
layer are forced by the liquid's dynamics). Given that, in the end, the two
regions will be matched, the boundary-layer scaling for $w$ can be deduced
from the small-$\bar{\rho}$ asymptotics of the liquid-region solution
(\ref{B.36}), which suggests%
\[
w_{i}=\rho_{v}w.
\]
As for the horizontal velocity $u$, the liquid-region solution (\ref{B.35})
implies that $u$ remains order-one when $\bar{\rho}\rightarrow0$ -- hence, in
the boundary layer, $u$ does not need to be rescaled.

One can also take advantage of two physical assumptions. Since the fluid
density in the boundary layer is small, one can safely assume that the
chemical potential there is that of ideal gas,%
\[
G(\rho,T)=T\ln\rho\qquad\text{if}\qquad\rho\sim\rho_{v}.
\]
In addition, both kinetic theory (e.g., \cite{FerzigerKaper72}) and
measurements (e.g., \cite{LindstromMallard97}) suggest that the vapor
viscosity and thermal conductivity are independent of the density -- thus, to
leading order, one can assume\begin{widetext}%
\[
\mu_{s}(\rho,T)=\mu_{s.v}(T),\qquad\mu_{b}(\rho,T)=\mu_{b.v}(T),\qquad
\kappa(\rho,T)=\kappa_{v}(T)\qquad\text{if}\qquad\rho\sim\rho_{v}.
\]
\end{widetext}Note that the viscosity and thermal conductivity of vapor are
typically much smaller than those of liquid -- hence, $\mu_{s.v}$, $\mu_{b.v}%
$, and $\kappa_{v}$ are small parameters (in addition to $\varepsilon$ and
$\rho_{v}$).

The following asymptotic limit is assumed:%
\begin{equation}
\varepsilon^{2}\rho_{v}^{-5/2}\mu_{v}=\mathcal{O}(1)\qquad\text{as}%
\qquad\varepsilon,\rho_{v},\mu_{v},\kappa_{v}\rightarrow0, \label{B.37}%
\end{equation}
where $\mu_{v}$ is, say, $(\mu_{s.v}+\mu_{b.v})/2$. As seen later,
(\ref{B.37}) is a characteristic limit of Regime 2 and, thus, covers adjacent
situations, $\varepsilon^{2}\rho_{v}^{-5/2}\mu_{v}\gg1$ and $\varepsilon
^{2}\rho_{v}^{-5/2}\mu_{v}\ll1$, as well. Note also that, as suggested by
measurements \cite{LindstromMallard97},%
\[
\kappa_{v}\sim\mu_{v}\gg\rho_{v}.
\]
Summarizing the above estimates, assumptions, and scaling, one can deduce from
the temperature equation (\ref{B.31}) the following quasi-isothermality
condition:%
\[
T=T_{0}+\mathcal{O}\left(  \varepsilon^{4}\frac{\rho_{v}^{3}}{\mu_{v}%
^{2}\kappa_{v}}\right)  ,
\]
which is even stronger than its liquid-region counterpart (\ref{B.31}).

Rewriting Eqs. (\ref{B.27}) and (\ref{B.29}) in terms of the new variables,
omitting Eq. (\ref{B.28}) for $u$ (which will not be needed), and replacing
the temperature equation with $T=\operatorname{const}$, one obtains
\begin{equation}
\frac{\partial\left(  \rho_{i}w_{i}\right)  }{\partial z_{i}}=\mathcal{O}%
(\rho_{v}^{3/2}), \label{B.38}%
\end{equation}%
\begin{multline}
\frac{\partial}{\partial z_{i}}\left[  T\ln\rho_{i}-\varepsilon^{2}\left(
\frac{\partial Z}{\partial x}\right)  ^{2}\frac{\partial^{2}\rho_{i}}{\partial
z_{i}^{2}}-\frac{\partial^{2}\rho_{i}}{\partial z_{i}^{2}}\right] \\
=\varepsilon^{2}\left[  \frac{\varepsilon^{2}\rho_{v}^{-5/2}\left(  \mu
_{b.v}+\frac{4}{3}\mu_{s.v}\right)  }{\rho_{i}}\frac{\partial^{2}w_{i}%
}{\partial z_{i}^{2}}+\mathcal{O}(\rho_{v}^{3/2})\right]  . \label{B.39}%
\end{multline}
Observe that $w_{i}$ appears in Eq. (\ref{B.39}) only as a perturbation, which
is why Eq. (\ref{B.38}) includes the leading-order term only.

Eq. (\ref{B.38}) yields $\rho_{i}w_{i}=\operatorname{const}$, where the
constant can be determined by matching $w_{i}$ to the small-$\bar{\rho}$ limit
of the liquid-region solution (\ref{B.36}). Keeping in mind that $\hat{\rho
}(z)$ is also small [because $\bar{\rho}(z)$ is small -- see (\ref{4.4})], one
obtains%
\begin{equation}
w=-\frac{\rho_{l}}{\bar{\rho}}\left\{  \frac{\partial h}{\partial t}-\rho
_{l}\frac{\partial}{\partial x}\left[  \frac{\partial G_{0}}{\partial
x}Q(h)\right]  \right\}  . \label{B.40}%
\end{equation}
$\rho_{i}$ should be sought in the form of a two-term expansion,%
\begin{equation}
\rho_{i}=\bar{\rho}_{i}(z_{i})+\varepsilon^{2}\rho_{i}^{(2)}+\cdots,
\label{B.41}%
\end{equation}
where $\bar{\rho}_{i}(z_{i})$ satisfies the following boundary-value problem%
\begin{equation}
\frac{\mathrm{d}^{2}\rho_{i}}{\mathrm{d}z_{i}^{2}}-T\ln\bar{\rho}_{i}=0,
\label{B.42}%
\end{equation}%
\begin{align}
\bar{\rho}  &  \rightarrow\infty\qquad\text{as}\qquad z^{\prime}%
\rightarrow-\infty,\label{B.43}\\
\bar{\rho}  &  \rightarrow1~\,\qquad\text{as}\qquad z^{\prime}\rightarrow
+\infty. \label{B.44}%
\end{align}
Physically, $\bar{\rho}_{i}$ describes the small-density part of a flat
interface in an unbounded space -- hence, it is the small-$\rho_{v}$ limit of
the function $\bar{\rho}$ defined previously.

Substitution of (\ref{B.41}) into Eq. (\ref{B.39}) yields%
\begin{multline}
\frac{\partial^{2}\rho_{i}^{(2)}}{\partial z_{i}^{2}}-\frac{T}{\bar{\rho}_{i}%
}\rho_{i}^{(2)}=-\left(  \frac{\partial Z}{\partial x}\right)  ^{2}%
\frac{\partial^{2}\bar{\rho}_{i}}{\partial z_{i}^{2}}\\
+\int_{z_{i}}^{\infty}\frac{\varepsilon^{2}\rho_{v}^{-5/2}\left(  \mu
_{b.v}+\frac{4}{3}\mu_{s.v}\right)  }{\rho_{i}}\frac{\partial^{2}w_{i}%
}{\partial z_{i}^{2}}\mathrm{d}z_{i}. \label{B.45}%
\end{multline}
The solution of this equation, $\rho_{i}^{(2)}$, should be matched to its
liquid-region counterpart $\rho^{(2)}$ (the first terms in the two expansions
match automatically, as they both describe a flat interface in an unbounded
space). Instead of $\rho_{i}^{(2)}$ and $\rho^{(2)}$, however, it is much
simpler to match the right-hand sides of the equations determining them:
(\ref{B.45}) for $\rho_{i}^{(2)}$ and (\ref{B.33}) for $\rho^{(2)}$ (the
left-hand sides of these equations match automatically under the assumption
that $G(\rho,T)\sim T\ln\rho$ as $\rho\rightarrow0$). Keeping in mind that
that $w_{i}$ is given by expression (\ref{B.40}), one obtains%
\[
\left(  \frac{\partial Z}{\partial x}\right)  ^{2}=\left[  \frac
{\partial(H+h)}{\partial x}\right]  ^{2},
\]
\begin{widetext}%
\begin{equation}
-G_{0}=\varepsilon^{2}\rho_{v}^{-5/2}\left(  \mu_{b.v}+\frac{4\mu_{s.v}}%
{3}\right)  \left\{  \frac{\partial h}{\partial t}-\rho_{l}\frac{\partial
}{\partial x}\left[  \frac{\partial G_{0}}{\partial x}Q(h)\right]  \right\}
\int_{-\infty}^{\infty}\frac{1}{\bar{\rho}_{i}}\frac{\partial^{2}}{\partial
z_{i}^{2}}\left(  -\frac{1}{\bar{\rho}_{i}}\right)  \mathrm{d}z_{i}%
.\label{B.46}%
\end{equation}
The former equality implies $Z=H+h+\operatorname{const}$, which means that the
boundary layer is pinned to a certain point of the interfacial profile. This
point is determined by $\operatorname{const}$, which can be found only from
the next order of the perturbation expansion (and does not affect the
leading-order solution).
Finally, using (\ref{B.34}) to eliminate $G_{0}$ from Eq. (\ref{B.46}),
one obtains the desired equation for $h(x,t)$ -- which can be written in form
(\ref{4.1}) with%
\begin{equation}
A=\rho_{l}^{2}\rho_{v}^{-5/2}\left(  \mu_{b.v}+\frac{4}{3}\mu_{s.v}\right)
\int_{-\infty}^{\infty}\frac{1}{\bar{\rho}_{i}^{4}}\left(  \frac
{\mathrm{d}\bar{\rho}_{i}}{\mathrm{d}z_{i}}\right)  ^{2}\mathrm{d}z_{i},
\label{B.47}%
\end{equation}
where the notation \textquotedblleft$A$\textquotedblright\ is used because
this coefficient is the small-$\rho_{v}$ limit of the Regime-1 coefficient $A$
[see formulae (\ref{B.25}) and (\ref{B.15})].
It remains to transform expression (\ref{B.47}) into its more convenient
version (\ref{4.2}). This can be done by changing the variable of integration
$z_{i}\rightarrow\bar{\rho}_{i}(z_{i})$, where the latter satisfies
boundary-value problem (\ref{B.42})--(\ref{B.44}). Thus, one can deduce that%
\[
A=\left(  \mu_{b.v}+\frac{4}{3}\mu_{s.v}\right)  T^{1/2}\rho_{v}^{-5/2}%
\int_{1}^{\infty}\frac{\sqrt{2\left(  \bar{\rho}_{i}\ln\bar{\rho}_{i}%
+1-\bar{\rho}_{i}\right)  }}{\bar{\rho}_{i}^{4}}\mathrm{d}\bar{\rho}_{i}.
\]
Evaluating the integral in the above expression numerically, one obtains
(\ref{4.2}) as required.\end{widetext}

\bibliography{}

\end{document}